\documentclass[aps,prL,twocolumn,superscriptaddress,showpacs]{revtex4-1}
\usepackage{graphicx}
\usepackage[dvipsnames,usenames]{color}
\usepackage{amsmath} 
\usepackage{amssymb}
\usepackage{bm}
\usepackage{multirow}
\usepackage{wasysym}
\usepackage{subfigure}
\usepackage{verbatim}
\usepackage{marvosym}
\usepackage{latexsym}
\usepackage[dvipsnames,usenames]{color}
\usepackage{float} 
\usepackage{ulem}

\usepackage{xr}
\makeatletter
\newcommand*{\addFileDependency}[1]{
  \typeout{(#1)}
  \@addtofilelist{#1}
  \IfFileExists{#1}{}{\typeout{No file #1.}}
}
\makeatother

\newcommand*{\myexternaldocument}[1]{
    \externaldocument{#1}
    \addFileDependency{#1.tex}
    \addFileDependency{#1.aux}
}

\myexternaldocument{new-SM}

\begin{document}
\title{Interfacial fluid rheology of soft particles}

\author{Maximilian M. Schmidt}
\affiliation{Institute of Physical Chemistry, RWTH Aachen University, Landoltweg 2, 52074 Aachen, Germany}

\author{Jos\'e Ruiz-Franco}
\affiliation{Laboratory of Physical Chemistry and Soft Matter, Wageningen University \& Research, Stippeneng 4, 6708 WE Wageningen, The Netherlands}

\author{Steffen Bochenek}
\affiliation{Institute of Physical Chemistry, RWTH Aachen University, Landoltweg 2, 52074 Aachen, Germany}

\author{Fabrizio Camerin}
\affiliation{Soft Condensed Matter \& Biophysics, Debye Institute for Nanomaterials Science, Utrecht University, Princetonplein 1, 3584 CC Utrecht, The Netherlands}

\author{Emanuela Zaccarelli}
\email{emanuela.zaccarelli@cnr.it}
\affiliation{Italian National Research Council - Institute for Complex Systems (CNR-ISC), Sapienza University of Rome, Piazzale Aldo Moro 5, 00185 Rome, Italy}
\affiliation{Department of Physics, Sapienza University of Rome, Piazzale Aldo Moro 2, 00185 Rome, Italy}

\author{Andrea~Scotti}
\email{andrea.scotti@mau.se}
\affiliation{Department of Biomedical Science, Faculty of Health and Society, Malm\"o University, SE-205 06 Malm\"o, Sweden}

\date{\today}
\begin{abstract}

\textit{In situ} interfacial rheology and numerical simulations are used to investigate microgel monolayers in a wide range of packing fractions, $\zeta_{2D}$. 
The heterogeneous particle compressibility determines two flow regimes characterized by distinct master curves. 
To mimic the microgel architecture and reproduce experiments, an interaction potential combining a soft shoulder with the Hertzian model is introduced. 
In contrast to bulk conditions, the elastic moduli vary non-monotonically with $\zeta_{2D}$ at the interface,  confirming long-sought predictions of reentrant behavior for Hertzian-like systems.

\end{abstract}

\maketitle 

Although softness is a concept used in everyday life, its impact on the macroscopic response of a material is still far from being fully understood \cite{Sco22_review}. 
Colloidal suspensions are fundamental model systems to investigate phase transitions and the behavior of complex fluids \cite{Tra01, Dur09, Gas01}.
Polymer-based colloids, particularly microgels \cite{Pel86}, allow us to investigate the effect of particle compressibility \cite{Hou22} on fundamental problems such as crystallization \cite{Pel15, Sco16,Rui18}, glass transition \cite{Mat09, vdS17, Pel16, bergman2018new}, and flow of non-Newtonian fluids \cite{Hel07, Mee04, Clo00, RuizPRM}.
Microgels have also been employed to investigate crystallization in 2D \cite{Deu13, Ill17,thorneywork2017two,rey2017anisotropic}. In fact, they spontaneously adsorb at interfaces reducing the interfacial tension between two liquids, undergoing a strong deformation and enhanced stretching \cite{Cam20}. 
They can be considered a model system for soft particles with non-homogeneous compressibility, due to their underlying core-corona structure \cite{Sco19, Boc22}, allowing us to introduce particle softness in the study of the phase behavior and flow properties in the two-dimensional interfacial plane.

Several recent {\it ex situ} experimental observations have reported that microgels crystallize in hexagonal lattices and undergo a solid-to-solid transition with increasing concentration \cite{Rey16, Boc19}. 
The structure of the monolayer has also been addressed {\it in situ} \cite{Kuk23, Kaw23}, where important differences have been observed with respect to {\it ex situ} measurements. 
It thus appears fundamental to address the properties of the monolayer by  {\it in situ} techniques \cite{Via20,Tat23} and by appropriate theoretical modeling \cite{Cia21, Har21, Cam20}.

Here, poly({\it{N}}-iso\-propylacryl\-amide) (pNIPAM) microgels are confined at the oil-water interface and their flow properties are investigated {\it in situ}. 
To this aim, amplitude and frequency sweeps, as well as flow curves, are measured using a rheometer with a custom-made double wall-ring accessory and a purpose-built Langmuir trough \cite{Van10, Tat23}.
We measure both the storage modulus and the apparent yield stress of the monolayer and find that they undergo a non-monotonic variation with increasing generalised packing fraction. 
This is in contrast to what is observed in bulk \cite{Con19, Pel16, Sch10}, where higher particle density leads to a monotonic increase of these quantities \cite{Ike13,Pel16, Sco20_flow}.

Experiments are supported by computer simulations in equilibrium, based on a coarse-grained model where a soft shoulder complements the Hertzian interaction (SSH potential), capturing the main features of the monolayer phase behavior. 
This model is further used to simulate the flow of the monolayer, confirming the non-monotonicity of the yield stress with increasing particle density.
Finally, we relate the flow behavior to the different regimes of the compression isotherms and identify the onset of two different master curves.
Our results highlight the impact of particle softness on the flow of the monolayer and the importance of inhomogeneous compressibility, which gives rise to multiple and unusual rheological regimes in ultra-dense conditions. 

\begin{figure}[htbp!]
    \centering
    \includegraphics[width=0.49\textwidth]{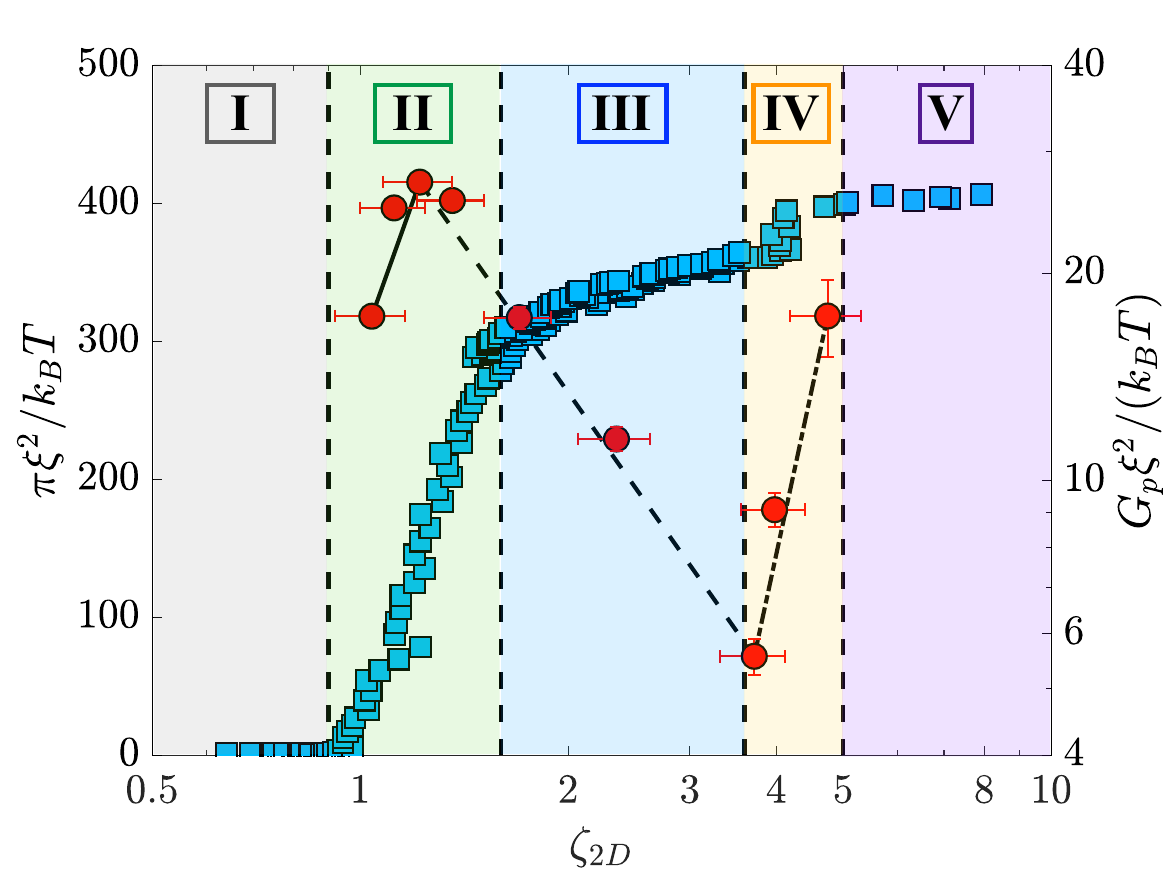}
    \caption{Compression isotherms reporting the surface pressure $\pi$ (squares), and plateau of the elastic modulus $G_p$ (circles), normalized by $k_BT/\xi^2$, as a function of generalised packing fraction, $\zeta_{2D}$. Lines are guides to the eye. Different colors identify different regimes of the compression isotherms.}
    \label{fig:Fig1}
\end{figure}

We perform experiments on 5~mol\% crosslinked microgels at an oil-water interface at 20~$^\circ$C.
In Fig.~\ref{fig:Fig1}, we report the compression isotherms of the microgels at the interface as a function of their interfacial generalised packing fraction, $\zeta_{2D}$, which corresponds to the area occupied by the microgels using their interfacial radius in the dilute limit, $R_{2D} \sim 350$ nm (Eq.~\ref{EQ_S1}).
This is estimated using atomic force microscopy (AFM) images of isolated microgels (Fig.~\ref{fig:R_dist})~\cite{Sco19, Sco22_review}.
The surface pressure, $\pi$,~vs.~$\zeta_{2D}$ behavior is consistent with previous measurements \cite{Gei14, Rey16, Sco22_review}, which have identified five regimes, corresponding to different compressions of the particles and to the formation of hexagonal lattices with different nearest neighbours distances (NND) \cite{Rey16}, see Figs.~\ref{fig:MB5Gr}-\ref{fig:MB5Psi6}.

To understand how the compression of the microgels affects the elasticity of the monolayer, \textit{in situ} oscillatory frequency sweeps were performed in the linear viscoelastic regime (Fig.~\ref{fig:MB5AS}) to measure the variation of the storage, $G^\prime$, and loss, $G^{\prime\prime}$, interfacial moduli (Fig.~\ref{fig:MB5_OFS}).
For all measured $\zeta_{2D}$, the monolayer can be considered a solid with $G^\prime > G^{\prime\prime}$. While $G^\prime$ remains almost constant, $G^{\prime\prime}$ shows a local minimum at a characteristic frequency, $\omega_m$. 
This behavior is consistent with the corresponding bulk measurements for microgel suspensions in the glassy state \cite{Pel16, Con19}. 
Conventionally, the plateau of the elastic modulus, $G_p$, is defined as  $G_p=G^\prime(\omega_m)$ \cite{Pel16}. 

In bulk (3D), $G_p^{3D}$ is usually normalized by the characteristic modulus associated with entropic elasticity, $k_BT/R_h^3$, where $k_B$ is the Boltzmann constant, $T$ the temperature and $R_h$ the hydrodynamic radius of the particles \cite{Pel16, Sco20_flow, Ike13}. 
Such a normalisation gives an idea of the relation between the internal energy of the system and the thermal fluctuations.
If $G_p^{3D}\gg k_BT/R_h^3$, thermal fluctuations are negligible and the physics determining the properties of the suspension is related to the particle deformation or interpenetration with its neighbour \cite{Con17, Con19, Sco19b, Hof22}.
At the interface, we can perform a similar normalisation using the ratio between $k_BT$ and the square of an appropriate characteristic length.
The particle radius is a natural first choice, but in the present study it always leads to $G_p\gg k_BT/R_h^2$, even at the lowest studied packing fraction, where the system is already a solid in the form of a hexagonal crystal \cite{Rey16}. 
This suggests that thermal fluctuations are always irrelevant for the monolayer and that its overall elastic properties are controlled  by length-scales much smaller than the particle size. Indeed, by considering the mesh size of the microgels ($\xi \approx 10$~nm)~\cite{Hou22} as the characteristic length, we obtain $G_p\xi^2/k_BT \approx 1$. 
This means that the elasticity of the monolayer is mainly determined by the capability of the particles to stretch at the interface, which is related to the deformability of the network, expressed by $\xi$~\cite{Sco22_review}.

In Fig.~\ref{fig:Fig1} (right y-axis), the normalized values of $G_p$ are plotted as a function of $\zeta_{2D}$.
In contrast to what observed in bulk \cite{Mat09, Con19, Sco16}, the evolution of $G_p$ with $\zeta_{2D}$ is non-monotonic. In particular, $G_p$ (circles) grows during regimes II and IV ($1.0\lesssim \zeta_{2D} \lesssim 1.6$ and $3.6 \lesssim \zeta_{2D} < 5.0$), while it decreases in regime III, in correspondence of the plateau of the measured compression isotherms (squares). 
Regimes II and IV correspond to the compression of regions of the microgels with different softness (external corona and core) \cite{Hou22, Sch21}.  
Fig~\ref{fig:MB5nnd} shows the values of the NND obtained from the analysis of AFM micrographs in dry state (Fig.~\ref{fig:MB5Gr}). 
The values of NND vary as $\zeta^{-1/2}$ (solid line) in regime II (circles), as expected for particles interacting \textit{via} soft repulsive potentials. 
To provide a microscopic interpretation of the non-monotonic behavior of $G_p$, we first note that in region II, the coronas of the microgels are still partially swollen,
but the compression of the monolayer progressively compacts them. 
However, until region III is reached, there is no microgel for which the corona is completely collapsed, with others still retaining a partial swollen corona (see AFM micrographs in Fig.~\ref{fig:MB5Gr}).
From the AFM images, it seems that particles with a collapsed corona form clusters, Fig.~\ref{fig:MB5Gr}.
However, recent studies have shown that the presence of clusters might be due to the monolayer deposition on a solid substrate \cite{Kuk23, Kaw23}.
Independently of the presence of the clusters, the key aspect that characterises regime III is that there are three different kinds of contact between microgels: (i) partially compressed corona-partially compressed corona; (ii) partially compressed corona-core; (iii) core-core (with potential presence of clusters). Thus, under these conditions, the monolayer has a more heterogeneous structure with core-core contacts continuously replacing corona-corona ones. This determines a less efficient transmission of the stress throughout the system, that we attribute to the fact that corona-corona contacts would ensure a higher degree of connectivity within the monolayer as compared to hard core contacts. This effect results in a decrease of $G_p$.
We further note that, if the clusters are present, they might also play a role in the decrease of $G_p$, similarly to what reported for depletion gels \cite{Con10, Hsi12, Whi19}. The following rise of $G_p$ in region IV is caused by a further increase of the packing fraction $\zeta_{2D}$ that brings the cores in closer and closer contact.

\begin{figure}[htbp!]
    \centering
    \includegraphics[width=0.49\textwidth]{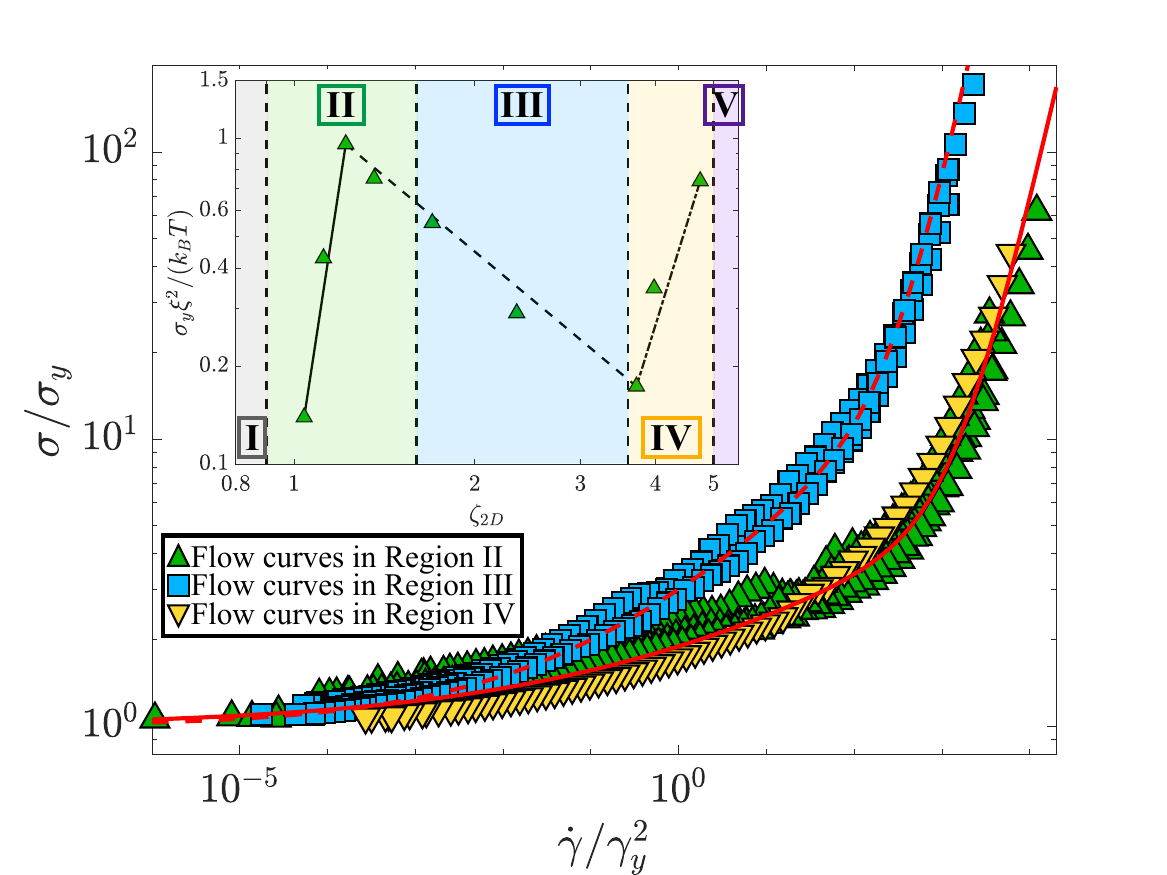}
    \caption{Master curves obtained by the flow curves measured in the different compression stages where the stress $\sigma$ is normalized by the apparent yield stress $\sigma_{y}$ and the shear rate $\dot{\gamma}$ is normalized by $\gamma_{y}^{2}$, where $\gamma_{y}^2$ is the threshold strain amplitude between liquid-like and solid-like behaviour.  
    Solid and dashed lines correspond to $\sigma/\sigma_y = 1 + k(\gamma/\gamma_y^2)^u + k'(\gamma/\gamma_y^2)^p$ with $p \simeq 0.2$ and $u \simeq 1.1$ and $p \simeq 0.3$ and $u \simeq 1.65$, respectively.
    Inset: $\sigma_y$, normalized by $k_BT/\xi^2$ as a function of the generalised area fraction, $\zeta_{2D}$. The different regimes of the compression isotherms in Fig.~\ref{fig:Fig1} are indicated by different colors.}
    \label{fig:Fig2}
\end{figure}

To probe the effect of softness on the flow of the microgel monolayer, we measured the flow curves at different $\zeta_{2D}$, Fig.~\ref{fig:MB5FC}.
These curves can be described by: $\sigma(\dot\gamma) = \sigma_y + k\dot\gamma^u + k'\dot\gamma^p$, where $\dot\gamma$ is the shear rate, $\sigma_y$ is named (apparent) yield stress by analogy to solids, $k$ and $k'$ are fitting parameters, $u$ and $p \in\mathbb{R}$ and are empirical parameters \cite{Erw10, Cag20}, see SM for more detail.
This is the same functional form used to fit flow curves of concentrated microgel suspensions in bulk \cite{Pel16, Sco20_flow}. 
In the inset of Fig.~\ref{fig:Fig2}, the triangles show the variation of $\sigma_y$ normalized by $k_BT/\xi^2$ as a function of $\zeta_{2D}$. 
The behavior of $\sigma_y$ is similar to that observed for $G_p$ in Fig.~\ref{fig:Fig1}(a), spanning an increase of roughly one order of magnitude upon initial compression, and undergoing a non-monotonic behavior at larger values of $\zeta_{2D}$. 
Interestingly, both moduli do not show an overall increase between the lowest and highest studied $\zeta_{2D}$, suggesting that within the explored range of packing fractions, the system does not reach a jammed state, according to the definition of Ref.~\cite{Pel16}.

After associating the different regimes of the compression isotherms with the non-monotonic response of $G_{p}$, we question whether this behavior affects the rheological response of the microgels at the interface under flow. To this aim, it is essential to define a unified scale that enables the evaluation of the flow behavior across different regimes and $\dot{\gamma}$. 
Following Cloitre and coworkers \cite{Pel16, Set11}, we plot $\sigma/\sigma_{y}$ vs.~$\dot{\gamma}/\gamma_{y}^{2}$, where $\gamma_{y}\propto\sigma_{y}/G_{p}$~\cite{Set11}.
The values of $\gamma_y$ and the corresponding errors have been determined by fitting the two different slopes of the flow curves measured in the amplitude sweeps (Fig.~\ref{fig:MB5AS}). 
As shown in Fig.~\ref{fig:gammay}(a), $\gamma_y$ rapidly increases in regime II before reaching a plateau in regime III, and then decreasing in regime IV.
Through this analysis, Fig.~\ref{fig:Fig2} shows two distinct master curves: the first one roughly corresponds to the monolayer in regimes II and IV  (triangles), while the second to regime III (squares) of the compression isotherm (Fig.~\ref{fig:MB5_HBparams}).
The normalisation by $1/\gamma_y^2$ is fundamental to obtain two well distinct master curves between regimes II/IV and III, as shown in Fig.~\ref{fig:gammay}(b).

\begin{figure*}[htbp!]
    \centering
    \includegraphics[width=\textwidth]{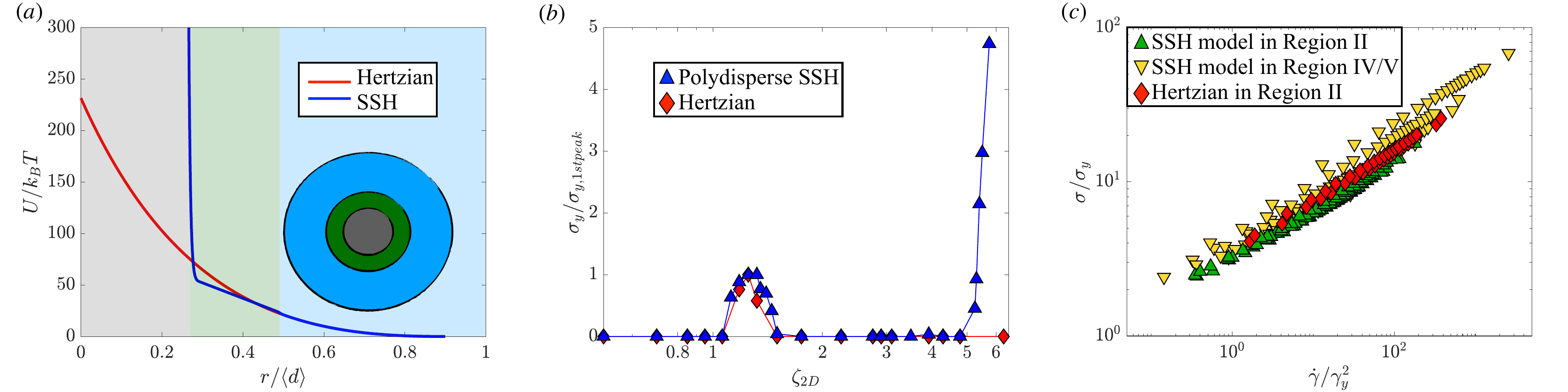}
    \caption{(a) Comparison between the 2D Hertzian potential (red curve) and the square-shoulder Hertzian model (SSH) (blue curve), where we highlight the three different regions of the particle: corona (light blue), compressible core (green) and incompressible core (gray). Corresponding colors are used in the background of the panel; 
    (b) apparent yield stress $\sigma_{y}$ normalized to the values of the first maximum of $\sigma_y$ ($\sigma_{y, 1st peak}$) as a function of $\zeta_{2D}$; (c) scaling flow curves,  $\sigma/\sigma_{y}$ versus $\dot{\gamma}/\gamma_{y}^2$, obtained from simulations using the monodisperse Hertzian (diamonds) and the polydisperse SSH (triangles) potential.
    }
    \label{fig:Fig3}
\end{figure*}

To shed light on the particle-to-particle interactions that give rise to this behavior, we introduce a novel interaction potential. 
This is needed since a simple 2D Hertzian model, shown in Fig.~\ref{fig:Fig3}(a), is too soft and fails to reproduce the behavior of the moduli at high $\zeta_{2D}$, as well as the evolution of NND with compression, see Fig.~\ref{fig:MB5nnd}. 
We thus complement the Hertzian, which works very well up to regime II, with a square shoulder potential \cite{sandoval2022soft, jagla1998phase}, mimicking the presence of a stiffer core, as illustrated in Fig.~\ref{fig:Fig3}(a). 
This also highlights the three characteristic lengths of the SSH model: the compressible corona, dominating at low and intermediate $\zeta_{2D}$, the standard core, e.g.~determined by the fuzzy sphere model where the square-shoulder takes over from the Hertzian, and, finally, an incompressible core, emerging at very large packing fractions due to the fact that the smallest value of the NND does not decrease any further. 
Simulations with the SSH model, also accounting for the experimental size polydispersity, are able to quantitatively reproduce the behavior of NND for both lengths, as a function of $\zeta_{2D}$, as shown in Figs.~\ref{fig:MB5nnd} and \ref{fig:gr_H} to \ref{fig:MSD}.

We then perform simulations of the monodisperse 2D Hertzian and of the polydisperse SSH models under steady shear (details in the SM).
For each studied shear rate $\dot{\gamma}$ and $\zeta_{2D}$, the mean value of the shear stress $\sigma_{xy}$ at the steady state is extracted. 
The resulting flow curves (Fig.~\ref{Sim_FC}) are well described by the standard Herschel-Bulkley model, i.e., $\sigma\left(\dot{\gamma}\right) = \sigma_{y}+k\dot{\gamma}^{u}$.
Fig.~\ref{fig:Fig3}(b) reports the yield stress $\sigma_{y}$ obtained from the fits of the simulated flow curves, normalized by the maximum of $\sigma_y$ for $\zeta_{2D} \sim 5.5$, that is the maximum packing fraction probed in the experiments.
For the Hertzian model, we observe a transition from a Newtonian fluid at low $\zeta_{2D}$ in regime I to a viscoelastic response with a non-monotonic behavior of $\sigma_{y}$ in regime II, before recovering a fluid behavior at higher area fractions. 
This reentrant behavior is a well-known hallmark for static and dynamic properties of Hertzian systems \cite{zhang2009thermal,Ber10,Cam20}, but to our knowledge it was not previously reported for elastic properties. 

For the SSH model, we find an almost identical qualitative behavior to the Hertzian up to regime III, that is then followed by a second growth of $\sigma_{y}$ in regime V.
The deviations between experiments and simulations in regimes III and IV (see Fig.~\ref{fig:Fig2}) are likely due to the coarse-grained nature of the model, which does not incorporate key polymeric aspects of the microgels. 
Indeed, it can be seen also in the experiments that the difference between $G^\prime$ and $G^{\prime\prime}$ in regime III is smaller (see Figs.~\ref{fig:MB5AS} and \ref{fig:MB5_OFS}), but the system remains solid-like probably because of entanglements between the coronas, a feature that cannot be captured by a simple point-particle model such as the present one. 
This difference may also explain the fact that the growth of $\sigma_{y}$ only occurs in regime V, rather than already in regime IV as in experiments, confirming the overall tendency of the simulations to retain a more fluid-like behavior than the experiments. 
Nonetheless, the re-increase of $\sigma_{y}$ observed at larger $\zeta_{2D}$ is well-captured by the model, thanks to the presence of the inner core in the SSH model. 
The comparison of the behavior of $\sigma_{y}$ with that of the mean-squared displacement (Fig.~\ref{fig:MSD}) demonstrates that the non-monotonicity of the moduli is the manifestation of such reentrant behavior in the elastic properties of the system.

We also analyze the scaling of the flow curves in simulations, using the same normalisation applied to the experimental data in Fig.~\ref{fig:Fig2}.
The plot reported in Fig.~\ref{fig:Fig3}(c), although spanning a smaller range than experiments, indicates that data in regimes II and IV/V roughly belong to the same master curve.  
Therefore, the behavior of the flow curves and the variation of $\sigma_y$ with $\zeta_{2D}$ confirm that
regimes II and IV/V are similar in terms of their flow and elastic properties. The rescaled flow curves obtained using the Hertzian model are also reported in Fig.~\ref{fig:Fig3}(c), being almost indistinguishable from those of the SSH model up to regime II. This confirms that before the corona compression, the interaction between microgels can be approximated with a simple Hertzian model, as predicted from numerical calculations of the effective potentials in the dilute limit~\cite{Cam20}.

In summary, we reported interfacial rheology measurements of monolayers of soft microgels at an oil-water interface. 
The experiments are compared to numerical simulations where a Hertzian is combined with a square shoulder potential, based on the core-corona structure of the microgels \cite{Sch21}. 
We find the onset of a non-monotonic behavior of the elastic moduli with increasing packing fraction due to the Hertzian contribution that is dominant at not too large packing fractions. 
Hence, after an initial fluid behavior (regime I), the system becomes viscoelastic (regime II), but the moduli start to decrease again (regime III), until a second increase is observed at very large $\zeta_{2D}$ (regimes IV and V).  
Simulations with such interaction potential qualitatively capture these features, except for the solid-like character of regime III.

The non-monotonic dependence of the moduli on packing fraction can also be interpreted in terms of the energy of the system: when particles have comparable softness, either before the microgel coronas start to be significantly compressed, or after the microgels make core-core contacts, the monolayer stores energy with increasing $\zeta_{2D}$.
The similarity between regime II and IV/V is further supported by the behavior of their flow curves, which are reduced to the same master curve. However, a different behavior (and associated master curve) applies in regime III, where the monolayer dissipates energy. 
This coincides with the onset of a second characteristic nearest neighbour distance in the radial distribution functions, highlighted by a plateau in the compression isotherms and denoting the emergence of core-core interactions. 
In this region, the contacts between particles are of different kinds, being either corona-corona, corona-core or core-core.
Such heterogeneities in the monolayer structure, and the different types of contacts between particles with different softness, decrease the transmission of the stress through the interface leading to the observed decrease in $G_p$.
These findings indicate the ability of the microgels to decrease stress thanks to deformation, suggesting that they can effectively reduce the viscous dissipation associated with the compression of their volume, by extending in the aqueous sub-phase~\cite{Ger23}. 

The present work sheds light on the flow properties of monolayers of soft particle at high concentrations. 
The fact that the measured elastic moduli do not really exceed those measured at the onset of elasticity suggests that the system does not reach jamming in the investigated range of packing fractions. 
This can be attributed to the role of individual particle softness and to the additional degrees of freedom of the particles outside the interfacial plane, e.g.,~through protrusion in the aqueous phase \cite{Ger23}. 
This peculiar feature of the interfacial behavior of microgels calls for further \textit{in situ} investigations, for example by neutron reflectivity~\cite{Boc22} or AFM~\cite{Via20}. 
These would clarify, for instance, the role of clusters on the flow properties of the monolayer and possible analogies with depletion glasses \cite{Con10, Hsi12, Whi19}, or shed light on the forces leading to cluster formation with possible similarity to the structures formed by proteins and antibodies at interfaces \cite{Tei20,Woo23}. 

The supporting data for this Letter are openly available~\cite{data23}.

\begin{acknowledgments}
The authors thank M. Brugnoni for the microgel synthesis and W. Richtering, A. Fernandez-Nieves, and M. Cloitre for fruitful discussions. M. M. S., S. B., and A. S. thank the Deutsche Forschungsgemeinschaft within projects A3 and B8 of the SFB 985 Functional Microgels and Microgel Systems. E. Z. acknowledges support from EU MSCA Doctoral Network QLUSTER, Grant Agreement No. 101072964 and ICSC---Centro Nazionale di Ricerca in High Performance Computing, Big Data and Quantum Computing, funded by the
European Union---NextGenerationEU-PNRR, Missione 4 Componente 2 Investimento 1.4.
\end{acknowledgments}

\bibliographystyle{apsrev4-1}
\bibliography{References}

\clearpage

\onecolumngrid

\section*{Supplemental Material for \\ ``Interfacial fluid rheology of soft particles''}
\setcounter{equation}{0}
\setcounter{figure}{0}
\setcounter{table}{0}
\setcounter{section}{0}

\renewcommand\thefigure{S\arabic{figure}} 
\noindent

\section{Microgel synthesis}

The microgels used here are the very same as in Refs.~\cite{Sco19b, Hou22} and they have been prepared by standard precipitation polymerisation \cite{Pel86}.
Briefly, 8.4870~g of NIPAM, 0.6090~g of BIS, and 0.0560~g of SDS were dissolved in 495~mL of filtered double-distilled water. 
The solution was heated to 60~$^\circ$C and degassed by purging with nitrogen for one hour under constant stirring at 300~rpm. 
The initiator solution, consisting of 0.2108~g of potassium persulfate (KPS) was diluted in 5~mL of filtered double-distilled water and degassed separately. 
The initiator was then transferred rapidly into the monomer solution to initiate the polymerisation. 
After 4~hours, the temperature was decreased to room temperature to stop the reaction.
The sample was then purified using ultra-centrifugation at 30,000~rpm and lyophilization was applied for storage.

\section{Microgel suspension preparation}

Microgel suspensions were prepared at a concentration of 0.25~wt\% by redispersing an appropriate amount of lyophilized microgel powder in the necessary volume of ultrapure water with 18.2~M$\Omega cm^{-1}$ resistivity. 
All suspensions were left on a rolling mixer overnight for equilibration.

\section{Generalised packing fraction}\label{sec:z_2d}

\begin{figure}[htb!]
    \centering
    \includegraphics[width=0.49\textwidth]{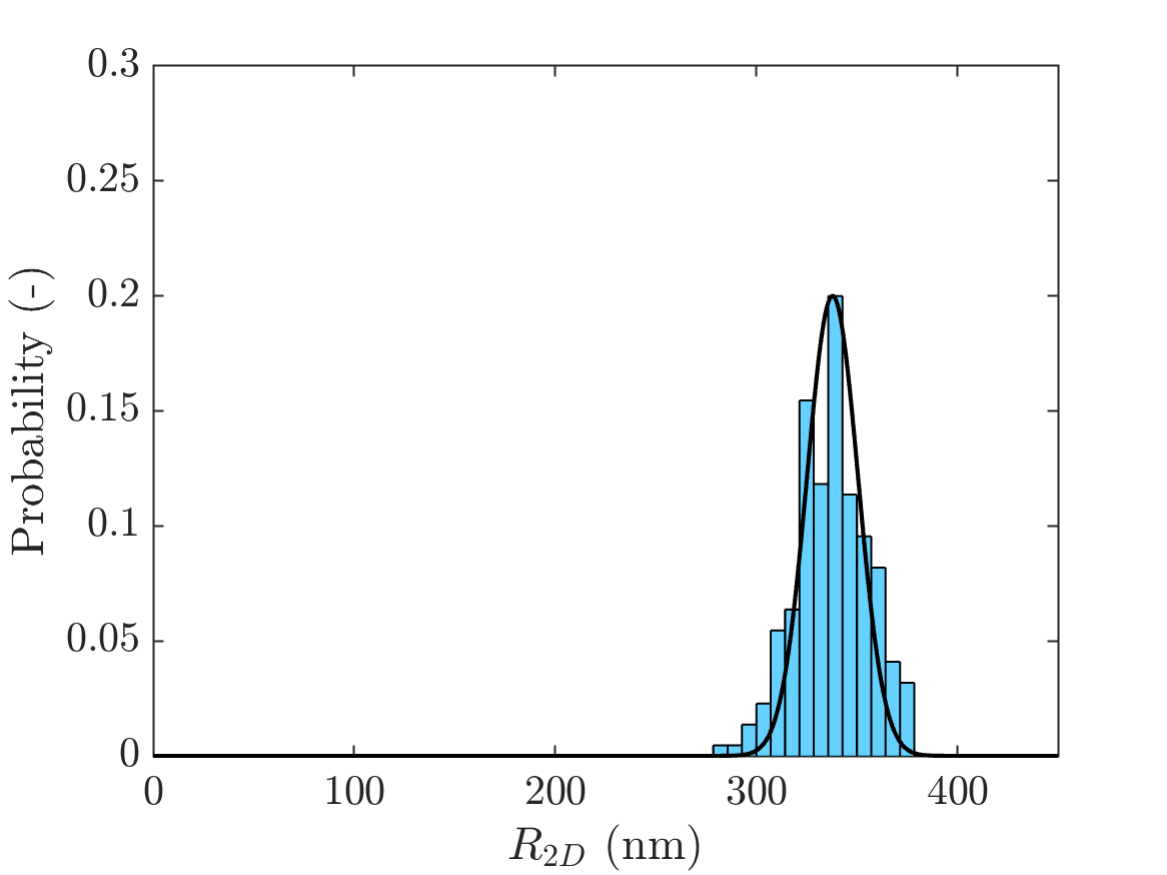}
    \caption{Distribution of the microgels radii determined from AFM images.}
    \label{fig:R_dist}
\end{figure}

The generalized packing fraction of microgels at the interface is defined as:

\begin{equation}
    \zeta_{2D} = \frac{N_p A_p}{A_{tot}}\,\, ,
    \label{EQ_S1}
\end{equation}

\noindent where $N_p$ is the number of localized microgel centers in an Atomic Force Miscropscopy (AFM) image, $A_{tot}$ is the total area of an image, and $A_p = \pi R_{2D}^2$ is the average area occupied by a particle of radius $R_{2D}$.
As particle radius, we consider the average radius of the adsorbed microgels at the interface before contact.
The radial distribution for the microgel used is shown in Figure~\ref{fig:R_dist} and is obtained considering $\gtrsim 220$ microgels.

\subsection{Estimation of the errors on $\zeta_{2D}$}

The errors on the $\zeta_{2D}$-values are calculated via Gaussian error propagation. 
However, in equation \ref{EQ_S1}, we only consider the values of $A_p$ to feature an error and those of $N_p$ and $A_{tot}$ to be without error. 
The error of the $A_p$ values is also determined via Gaussian error propagation taking into account the error of the $R_{2D}$ values. 
The latter corresponds to the standard deviation of the respective $R_{2D}$ value.

\section{Phase behavior at the interface}

\begin{figure*}[htbp!]
    \centering
    \includegraphics[width=\textwidth]{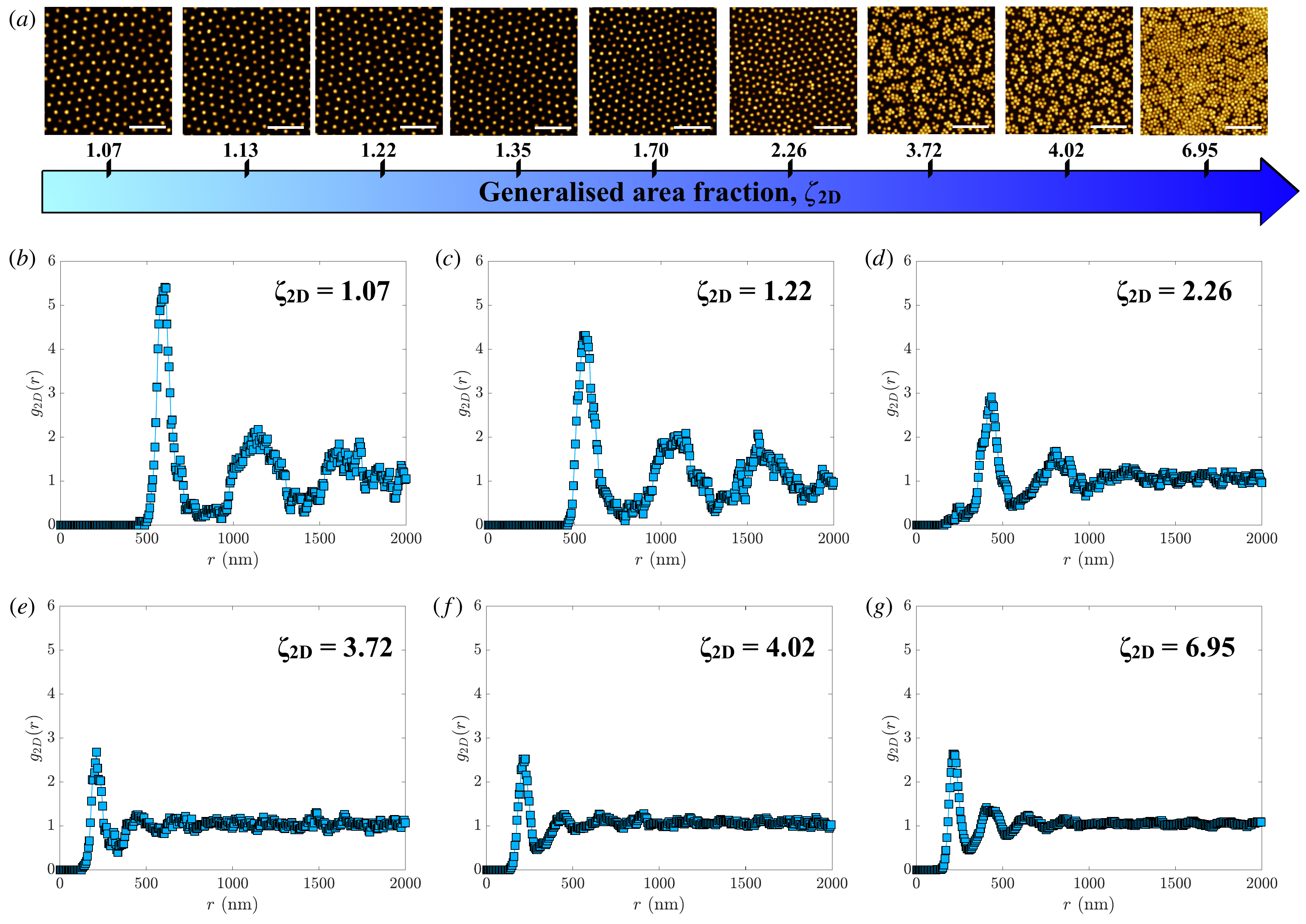}
    \caption{(a) Atomic force microscopy images of the monolayer of microgels deposited on a solid substrate. The generalized packing fraction increases from left to right. Scale bar is 2~$\mu$m. (b)-(i) Radial distribution functions $g_{2D}(r)$ corresponding to the images in (a) at $\zeta_{2D}$ equal $1.07\pm0.12$ (b), $1.22\pm0.14$ (c), $2.26\pm0.25$ (d) $3.72\pm0.40$ (e), $4.02\pm0.43$ (f) and $6.95\pm0.75$ (g).}
    \label{fig:MB5Gr}
\end{figure*}

Figure~\ref{fig:MB5Gr}(a) shows the AFM micrograph of the dry deposited microgel monolayer at increasing generalized packing fraction, $\zeta_{2D}$.

\begin{figure}[htbp!]
   \centering
   \includegraphics[width=0.5\textwidth]{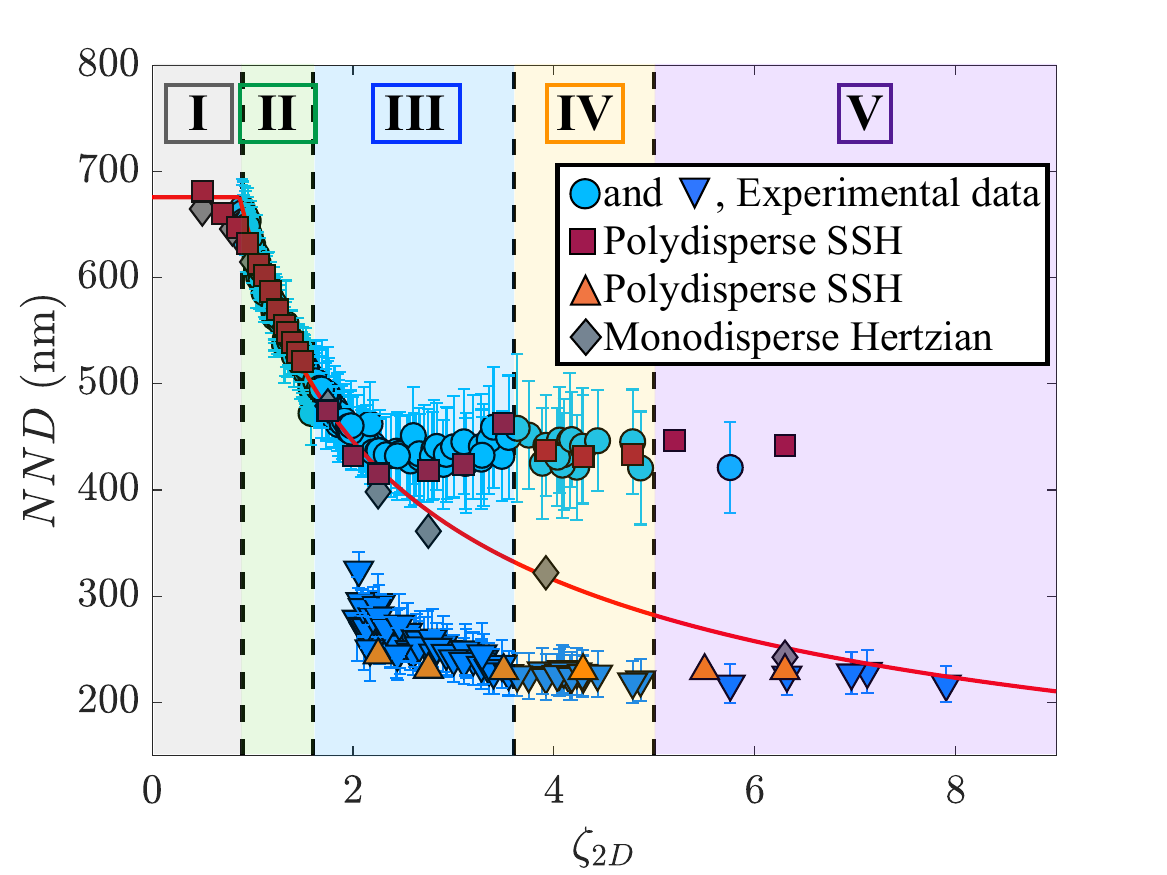}
   \caption{Nearest neighbours distance, $NND$, as a function of $\zeta_{2D}$. Circles and triangles represent the experimental values of the $NND$ before and after microgels get in close contact. 
   The $NND$ obtained from the simulated $g(r)$ are represented by diamonds (monodisperse Hertzian potential), squares (polydisperse SSH in first regime), triangles (polydisperse SSH in second regime). The solid line is a fit with $NND \simeq \zeta^{-1/2}$. The vertical lines mark the transitions between the different regimes,as deduced from the compression isotherm in Fig.~\ref{fig:Fig1}.}
   \label{fig:MB5nnd}
\end{figure}

As can be seen by the radial distribution function, $g_{2D}(r)$, the microgels form a hexagonal lattice characterized by structural peaks, Fig.~\ref{fig:MB5Gr}(b).
With increasing $\zeta_{2D}$, the first structural peak shifts to smaller values of the radius as a consequence of the  compression of the external corona that is the softest part of the microgels \cite{Hou22, Sch21}. 
From the analysis of the AFM images, the values of the nearest neighbour distance are extracted and plotted as a function of $\zeta_{2D}$, as shown by the circles in Fig.~\ref{fig:MB5nnd}. 
At low packing fractions, the distance follows the scaling associated to an isotropic compression.

At a critical $\zeta_{2D} \gtrsim 2.00$, the AFM micrographs show the presence of clusters, Fig.~\ref{fig:MB5Gr}.
Independently of the real presence of these structures at the interface \cite{Kuk23, Kaw23}, the important aspect is that in regime III we have different contacts between microgels, characterised by different softness: (i) partially compressed corona - partially compressed corona; (ii) partially compressed corona - core; (iii) core-core.
As a consequence of these different contacts, and corresponding different particle-to-particle distances, the $g_{2D}(r)$s corresponding to these images display a second structural peak at $r \approx 245$~nm, down-side triangles in Fig.~\ref{fig:MB5nnd}.
With further increase of $\zeta_{2D}$, the number of particle in core-core contact grows until all the particles are in this situation. 
The $g_{2D}(r)$'s are now consistent with a hexagonal lattice characterized by a smaller lattice constant, Figs.~\ref{fig:MB5Gr}(e)-to-\ref{fig:MB5Gr}(g).

The presence of particle clusters has also been observed in depletion gels composed of PMMA rigid particles and polystyrene polymer in bulk \cite{Con10, Hsi12, Whi19}.
We note that, despite the amount and number of clusters present in region III is still under debate \cite{Kuk23, Kaw23}, they might also play a role in the decrease of $G_p$. However, we should look at the similarity with depletion glasses with caution.
For instance, in contrast to depletion glasses where clusters forms from a fluid phase \cite{Con10}, here they form starting from a crystalline lattice.

\begin{figure}[htbp!]
    \centering
    \includegraphics[width=0.5\textwidth]{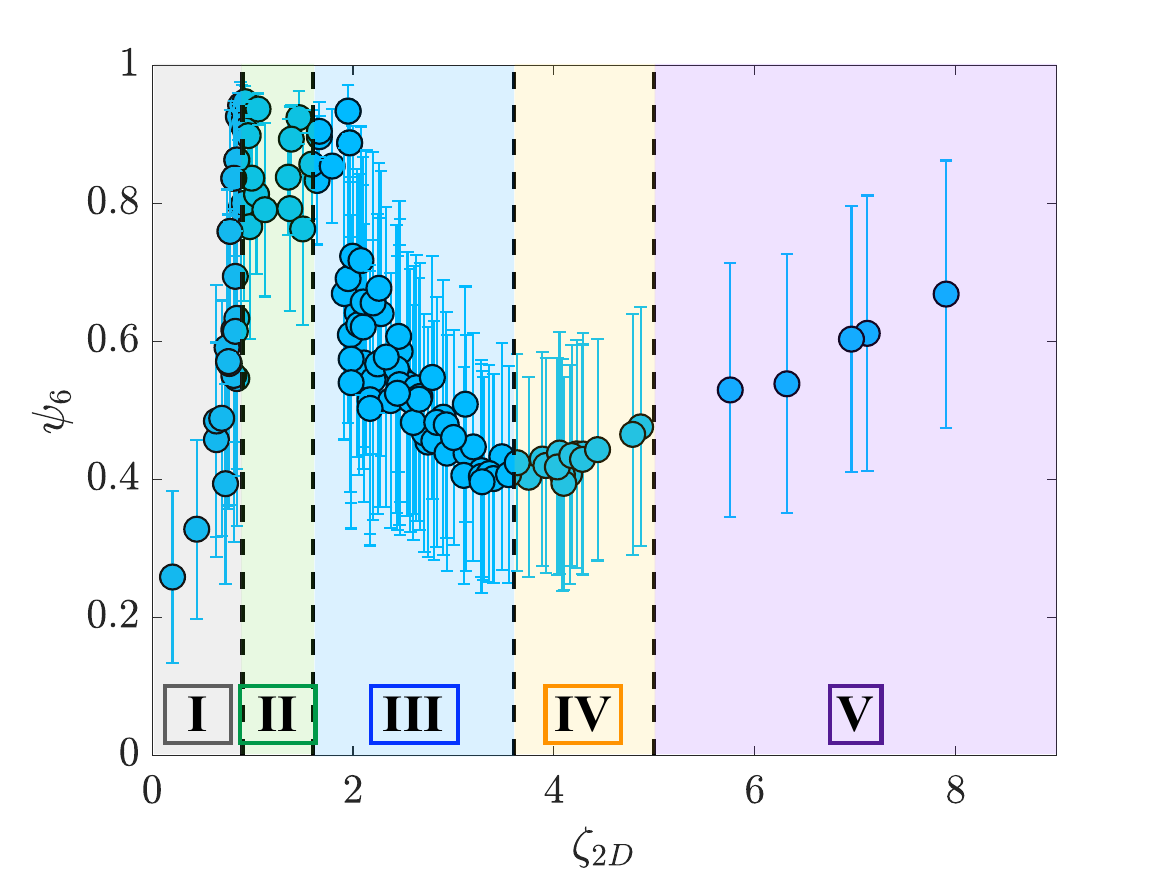}
    \caption{Order parameter $\psi_6$ vs.~generalised packing fraction of microgels in the monolayer, $\zeta_{2D}$.}
    \label{fig:MB5Psi6}
\end{figure}

To highlight the difference with depletion glasses and support the fact that the microgels in the monolayer form a crystalline lattice already in region II, before the appearance of the clusters, Fig.~\ref{fig:MB5Psi6} shows the so called order parameter $\psi_6$, calculated from the AFM micrographs in Fig.~\ref{fig:MB5Gr} \cite{Rey16, Sco19, Boc19}.
As can be seen, in region II, $\psi_6 \approx 1$, indicating the onset of short-range order. 
This, together with the $g(r)$'s, reported in Fig.~\ref{fig:MB5Gr}, shows that in this region we are dealing  with a crystalline lattice.
The decrease of $\psi_6$ and the progressive disappearance of the crystalline peaks in the $g(r)$ then tell us that in the occurrence of the isostructural transition, the first lattice melts while microgels with compressed corona get in contact.
This is also confirmed by the fact that the nearest neighbour distance has a discontinuous change between region II and IV, as shown in Fig.~\ref{fig:MB5nnd}.

\section{Interfacial shear rheology experiments}

Shear rheological experiments at oil-water interfaces were performed using a Discovery HR-3 rheometer (TA Instruments Inc., USA), combined with a custom-made double wall-ring (DWR) accessory and a purpose-built Langmuir trough \cite{Van10, Tat23}. 
The DWR accessory was based on the design introduced by \citeauthor{Van10} and utilized an open cup placed at the center of the trough.
The trough and the cup were both milled from polyoxymethylene (POM). 
The Langmuir trough was equipped with two customized movable barriers, also made from POM, and connected to an external thermostat (Julabo F12, Julabo GmbH, Germany). 

With the barriers in place and the DWR cup situated at the trough's center, the maximum interfacial area was $\approx 583$~cm$^2$, while the minimum interfacial area was fixed at $\approx 98$~cm$^2$. 
The surface pressure was monitored by a platinum Wilhelmy plate (KN 0002, KSV NIMA/Biolin Scientific Oy, Finland) with a perimeter of 39.24~mm, which was attached to an electronic film balance. 
The plate was positioned left-side of the cup and oriented parallel to the trough's movable barriers.

All interfacial rheological experiments were conducted at oil-water interfaces at 20~$^\circ$C. 
Decane ($\leq 94$\%, Merck KGaA, Germany) filtered three times over basic aluminum oxide (90 standardized, Merck KGaA, Germany) was employed as the oil top phase because of its non-solvent properties for N-isopropylacrylamide (NIPAM)-based microgels. 
The aqueous subphase comprised water of ultrapure quality with a resistivity of 18.2~M$\Omega cm^{-1}$. 
Aqueous microgel stock solutions were prepared for all experiments. 
Before their use, a stock solution was mixed with 10~vol\% isopropyl alcohol ($\leq99.8$\%, Merck KGaA, Germany) to facilitate the spreading of the microgels.

Three different types of experiments were performed on the microgel monolayers at various controlled surface pressures: amplitude sweep measurements, frequency sweep measurements, and flow curves. 
The general procedure to prepare a microgel-laden interface before the start of any measurement was always the same. 
First, the trough, the movable barriers, and the cup were thoroughly cleaned and assembled. 
Then, the aqueous subphase was added to the trough and checked for interfacial-active impurities. 
At the same time, the ring geometry was cleaned and coupled to the rheometer. 
After calibration, the geometry was slowly lowered towards the air-water interface until the ring was pinned. 
The exact height position of the ring was determined by monitoring the normal force acting on the geometry on approach and accounting for the diamond-shaped cross-section of the ring. 
Next, the oil top phase was carefully added to the trough, and the resulting oil-water interface was again checked for interfacial-active contamination. 

With the setup ready for use, a respective microgel stock solution was added drop-wise to the interface with the help of a glass syringe. 
The microgel-covered interface was left to equilibrate for 60~min. 
Afterward, the interfacial layer was compressed to a specified surface pressure by closing the movable trough barriers symmetrically at a rate of 10~mm min$^{-1}$. 
Once the targeted surface pressure was reached, it was maintained for another 15~min, after which one of the following rheological measurement was started.
Regardless of the type of experiment, the surface pressure was kept constant throughout a respective measurement by allowing the Langmuir trough's movable barriers to adjust their position automatically if needed.

\subsection{Amplitude frequency sweeps}

Amplitude sweep measurements at controlled surface pressures were conducted with an angular frequency of 1.0~rad s$^{-1}$. 
For each measurement, the applied shear strain amplitude was varied from 0.5\% to 100\%.

\begin{figure*}[htbp!]
    \centering
    \includegraphics[width=\textwidth]{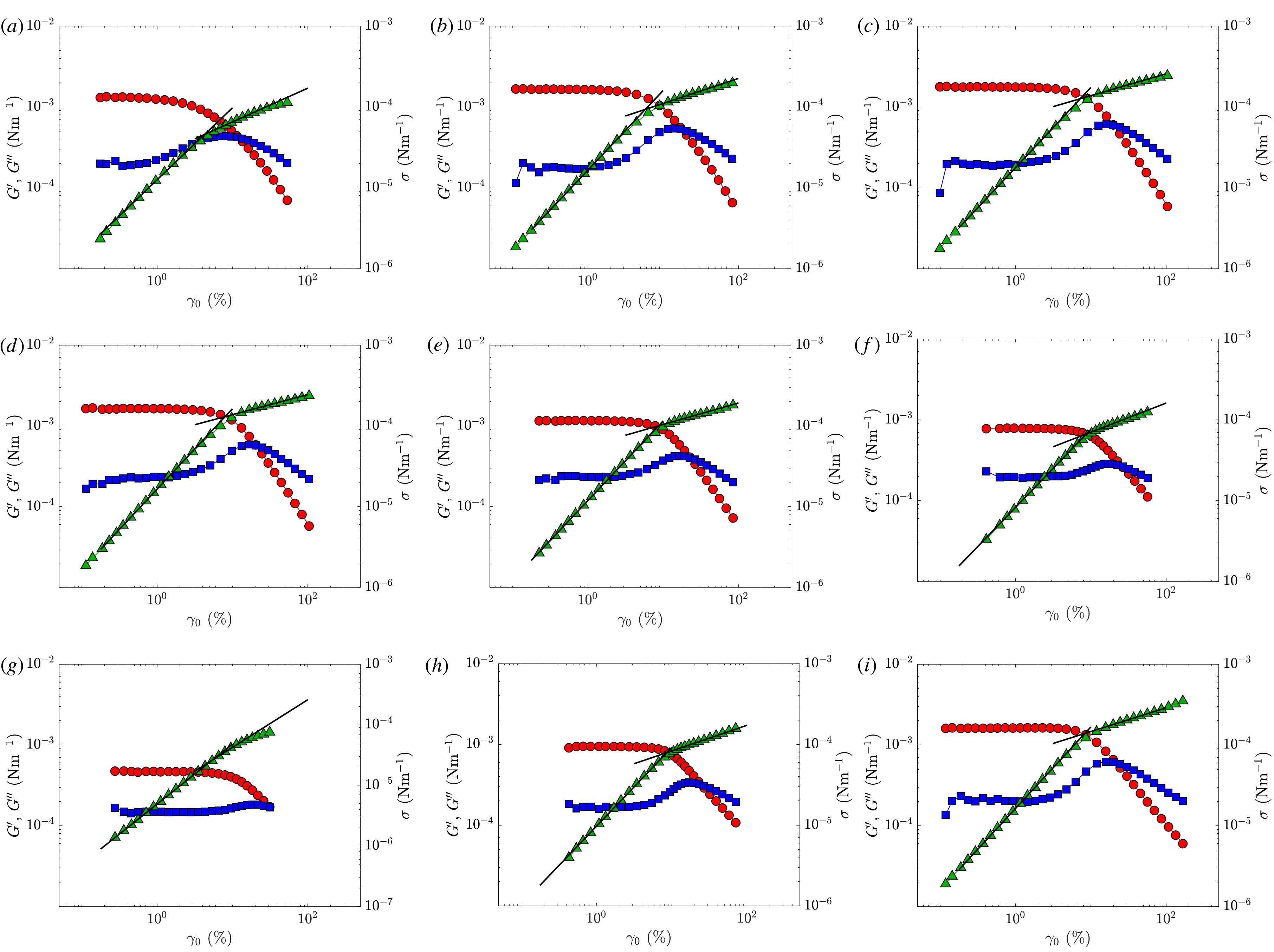}
    \caption{Values of the elastic (circles) and loss (squares) moduli, and stress (triangle, right axes) \emph{vs.}~strain, $\gamma_0$, measured in amplitude sweep ($\omega = 1$~rad s$^{-1}$) at different values of $\zeta_{2D}$: $1.04\pm0.12$ (a), $1.12\pm0.12$ (b), $1.22\pm0.14$ (c) $1.36\pm0.15$ (d), $1.70\pm0.19$ (e), $2.35\pm0.28$ (f), $3.72\pm0.40$ (g), $3.98\pm0.42$ (h), and $4.75\pm0.56$ (i). 
    Black solid lines are linear fits of $\sigma$~vs~$\gamma_0$ for high and low values of $\gamma_0$. 
    The intercept corresponds to $\gamma_y$.}
    \label{fig:MB5AS}
\end{figure*}

Figure~\ref{fig:MB5AS} shows the storage and loss moduli, $G^\prime$ (circles) and G$^{\prime\prime}$ (squares), and the stress amplitude, $\sigma$ (triangles), as a function of the strain amplitude, $\gamma_0$.
As in bulk, both the moduli of the monolayer are independent of the strain amplitude for low $\gamma_0$.
Also the stress behaves as expected and increases linearly for low $\gamma_0$.
All the measurements in Fig.~\ref{fig:MB5AS} ensure that at $\gamma_0 = 1\%$ ($\omega = 1$~rad s$^{-1}$), the monolayer can be considered in the linear viscoelastic regime.
With increasing $\gamma_0$, $G^{\prime}$ decreases.
In contrast, $G^{\prime\prime}$ increases, reaches a maximum and then decreases.
For these high $\gamma_0$, $\sigma$ still increases but with a different power law.
Conventionally, the value of $\gamma_0$, at which low- and high strain power law variations cross (solid lines), is considered to be the threshold strain amplitude between liquid-like and solid-like behaviour, $\gamma_y$.

\begin{figure*}[htbp!]
    \centering
    \includegraphics[width=0.49\textwidth]{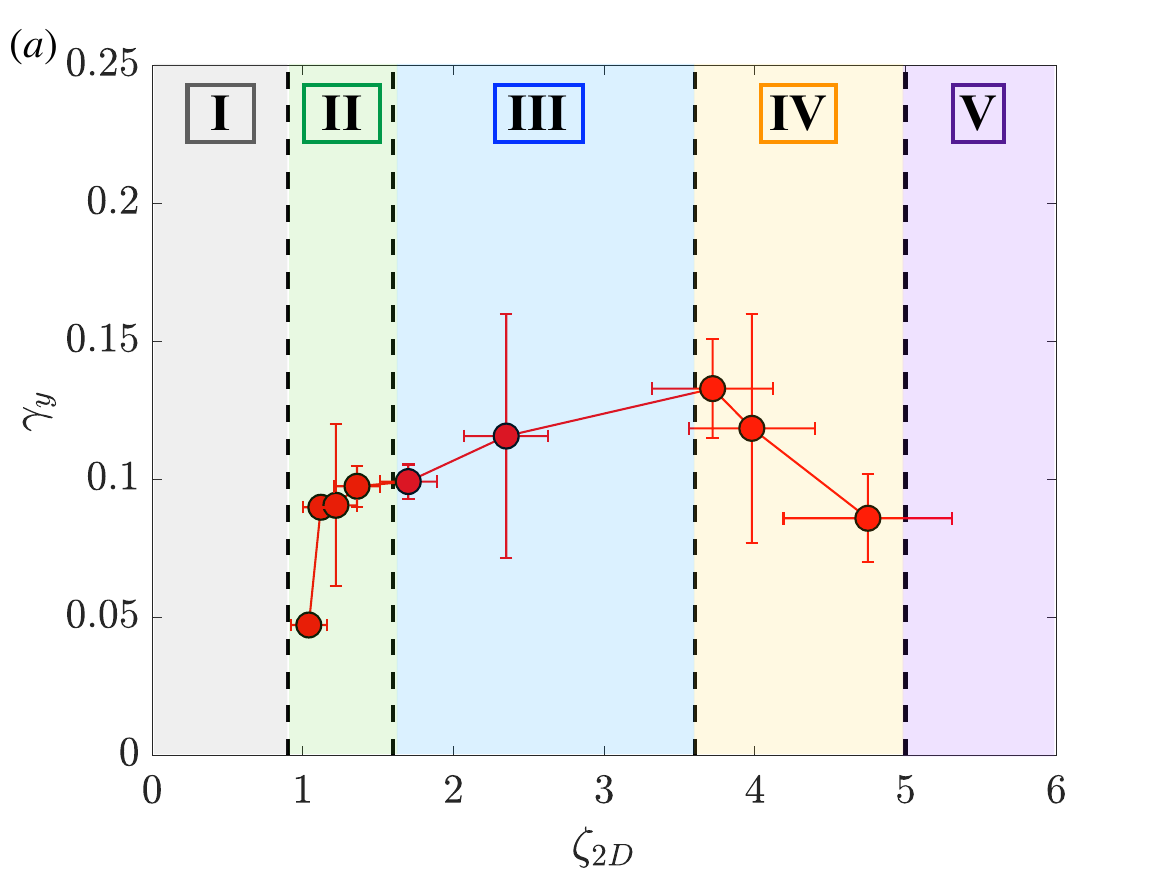}
    \includegraphics[width=0.49\textwidth]{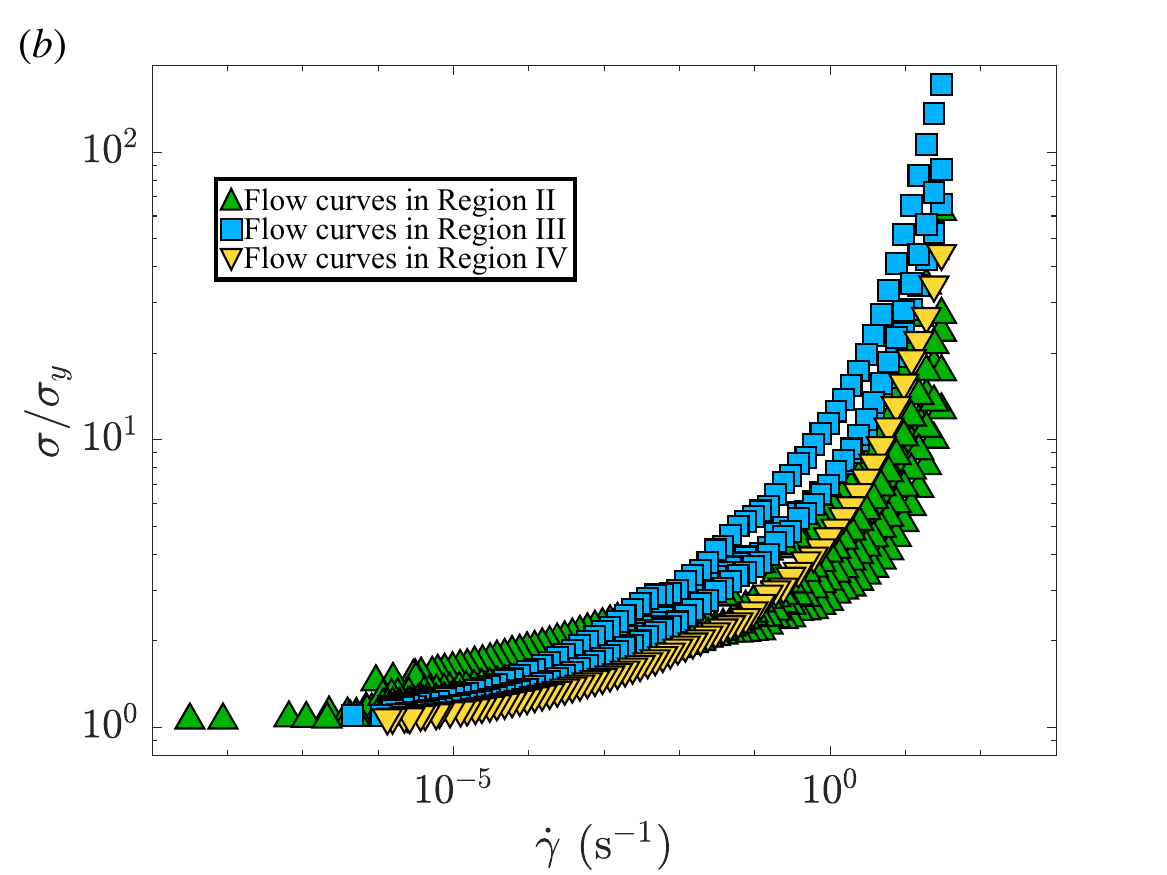}
    \caption{(a) Values of $\gamma_y$, as determined from the flow curves (triangles in Fig.~\ref{fig:MB5AS}), versus the packing fraction of the monolayer, $\zeta_{2D}$.
    (b) Flow curves measured in the different compression stages where the stress $\sigma$, normalised by the corresponding $\sigma_y$, is plotted as a function of the shear rate $\dot\gamma$.
    }
    \label{fig:gammay}
\end{figure*}

The values of $\gamma_y$ vary between 5 and 15\% as shown by the data in Fig.~\ref{fig:gammay}(a), with relative errors varying between 3 and 16\%. We find that $\gamma_y$ steeply increases in region II, then it keeps increasing smoothly in region III before decreasing in region IV.
The estimate of the errors is obtained using error propagation on the value of the intercept between the linear fits (solid black lines) of the flow curves (green triangles) in Fig.~\ref{fig:MB5AS}. 
If the normalisation performed in Fig.~\ref{fig:Fig2} is not applied, the flow curves do not show any common behavior, independently of the regime in which they have been measured, Fig.~\ref{fig:gammay}(b).

\subsection{Oscillatory frequency sweeps}

\begin{figure*}[htbp!]
    \centering
    \includegraphics[width=\textwidth]{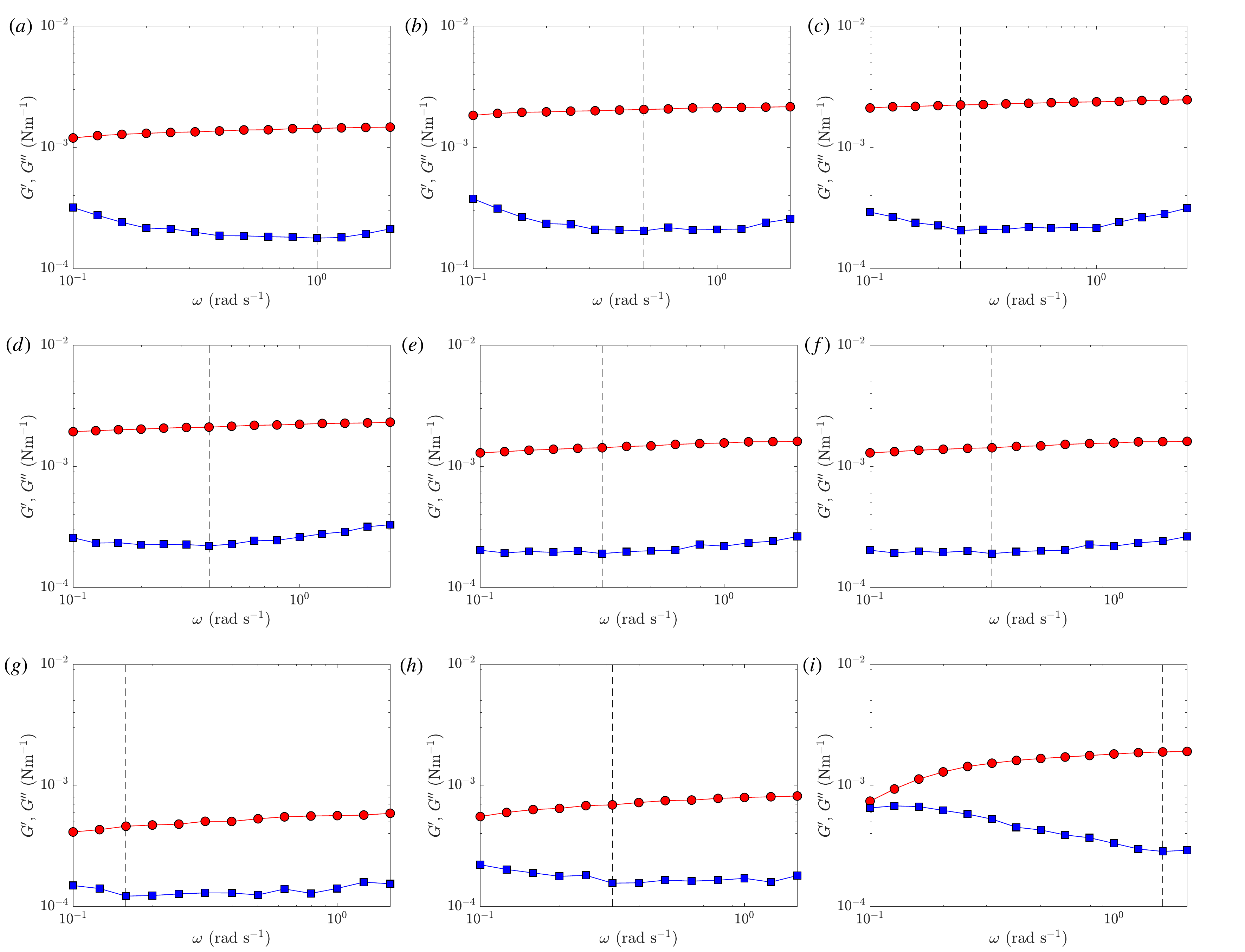}
    \caption{Values of the elastic (circles) and loss modulus (squares) \emph{vs.}~the angular frequency, $\omega$, measured in frequency sweep ($\gamma_0 = 1$\%) at different values of $\zeta_{2D}$: $1.04\pm0.12$ (a), $1.12\pm0.12$ (b), $1.22\pm0.14$ (c) $1.36\pm0.15$ (d), $1.70\pm0.19$ (e), $2.35\pm0.28$ (f), $3.72\pm0.40m$ (g), $3.98\pm0.42$ (h), and $4.75\pm0.56$ (i).}
    \label{fig:MB5_OFS}
\end{figure*}

Frequency sweep measurements at controlled surface pressures were carried out with a shear strain amplitude of 1\%, i.e.~in the linear viscoelastic region (see Figure~\ref{fig:MB5AS}). 
For each measurement reported in Figure~\ref{fig:MB5_OFS}, the applied angular frequency was varied from 0.1~rad s$^{-1}$ to 3.0~rad s$^{-1}$. 

\begin{figure*}[htbp!]
    \centering
    \includegraphics[width=\textwidth]{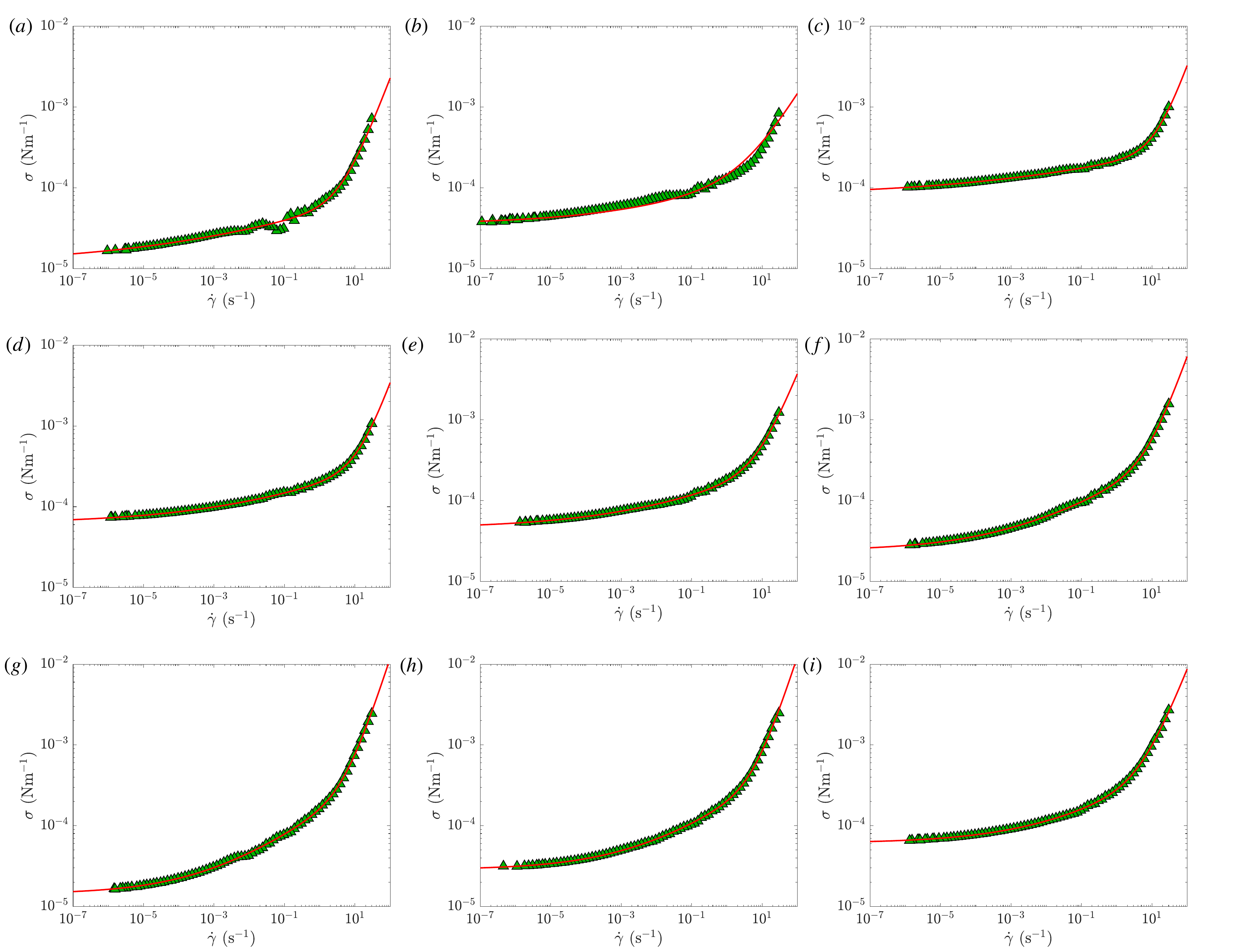}
    \caption{ Values of the shear stress (squares), $\sigma$, \emph{vs.}~the shear-rate, $\dot\gamma$, at different values of $\zeta_{2D}$: $1.04\pm0.12$ (a), $1.12\pm0.12$ (b), $1.22\pm0.14$ (c) $1.36\pm0.15$ (d), $1.70\pm0.19$ (e), $2.35\pm0.28$ (f), $3.72\pm0.40$ (g), $3.98\pm0.42$ (h), and $4.75\pm0.56$ (i). 
    The solid lines are fits with the Herschel–Buckley model, Eq.~\ref{eq:shear_vs_gdot}.}
    \label{fig:MB5FC}
\end{figure*}

\subsection{Flow curves}

To quantify the value of $\sigma_y$ the $\sigma$~vs.~$\dot\gamma$, data are fitted using a linear combination of a constant shear stress and two power laws:
\begin{equation}
\sigma(\dot\gamma) = \sigma_y + k\dot\gamma^u + k'\dot\gamma^p \label{eq:shear_vs_gdot}
\end{equation}
where $k$, $k'$ are fitting parameters, $u$ and $p \in\mathbb{R}$ and $\sigma_y$ is named (apparent) yield stress in analogy to solids. 
The second additional term with respect to the classic  Herschel-Bulkley equation, $k^\prime\sigma^p$, has been interpreted to account for how corona deformability contributes to the system's flow under deformation \cite{Erw10}.
Equation~\ref{eq:shear_vs_gdot} was previously used to fit $\sigma$~vs.~$\dot\gamma$ curves for suspensions of hard and soft colloids in bulk, and here it successfully reproduces the measured flow curves for the monolayer at different $\zeta_{2D}$, solid lines in Fig.~\ref{fig:MB5FC}.

\begin{figure}[htbp!]
    \centering
    \includegraphics[width=1\textwidth]{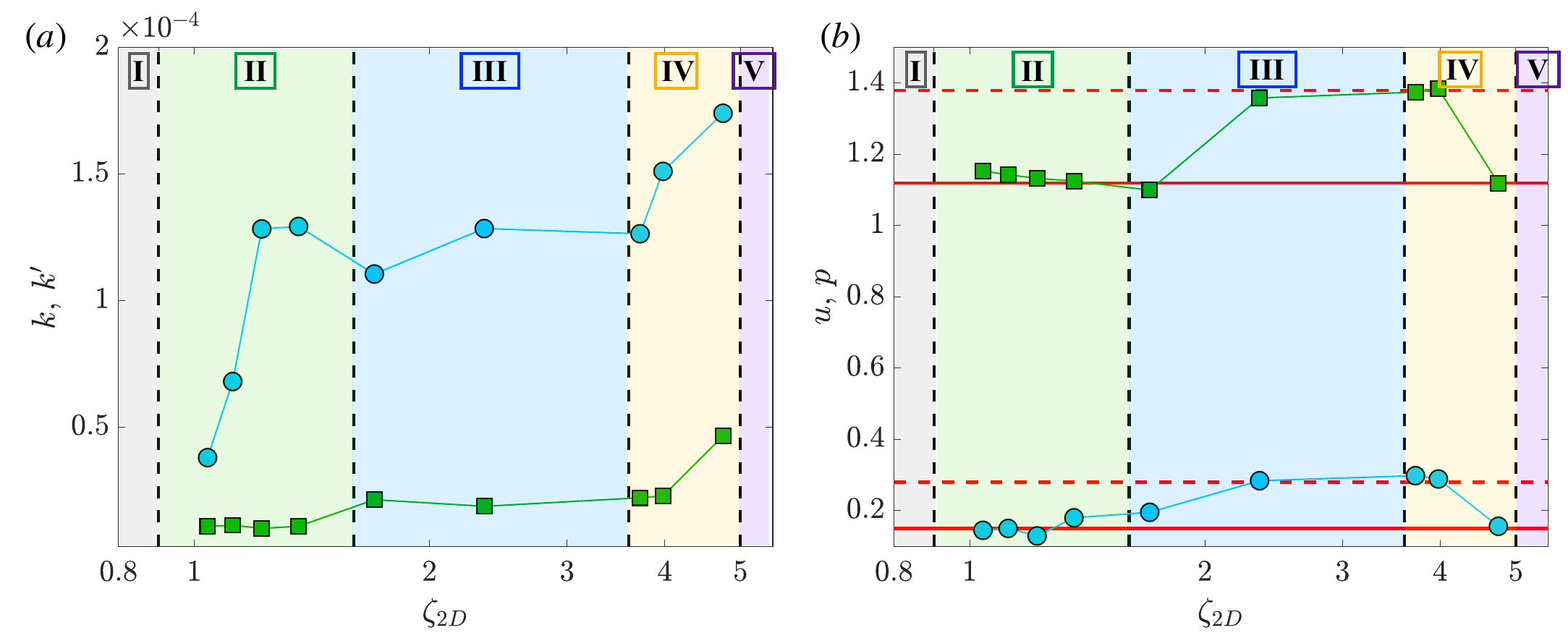}
    \caption{Parameters obtained from fitting the experimental flow curves with $\sigma/\sigma_y = 1 + k\tilde{\gamma}^u + k'\tilde{\gamma}^p$: (a) prefactors $k$ (circles) and $k^\prime$ (squares) and (b) exponents $u$ (squares) and $p$ (circles) as a function of $\zeta_{2D}$. 
    Vertical dashed lines mark the different regions of the compression isotherms identified in Fig.~\ref{fig:Fig1} of the manuscript. 
    Horizontal red lines in (b) mark the values of the exponents for the master curves shown in Fig.~\ref{fig:Fig2} of the manuscript: solid lines for Regions II and IV and dashed lines for Region III.}
    \label{fig:MB5_HBparams}
\end{figure}

\noindent Fig.~\ref{fig:MB5_HBparams} shows the evolution of the prefactors $k$ and $k'$ and of the exponents $u$ and $p$ obtained from the fits.
As  it can be seen from the figure, for the curves that belong to the second flow regime (region III) the values of the exponents change.
As mentioned in the manuscript, $u$ (squares) changes from $\simeq 1.1$ to $\simeq 1.4$ and $p$ (circles) changes from $\simeq0.15$ to $\simeq 0.3$.
In bulk, the values reported for the exponents are $0.4\lesssim u \lesssim 0.7$ and $0.15\lesssim p \lesssim 0.2$ \cite{Pel16}.
A value of $u \sim 0.5$ was observed in bulk experiments and interpreted to be associated to internal slip between localized randomly oriented planes of the particles flowing on top of each other, when the particle are deformables and the dynamics is solvent-mediated \cite{Set11}.
Given the fact that we are limited to an interface, it is not surprising that we do not find a similar value.

We also note that, for the present study, the ratio between prefactors is $0.08\lesssim k/k^\prime \lesssim 0.3$ that is comparable to the ratio between these two parameters observed in bulk $0.05\lesssim k/k^\prime \lesssim 0.3$.
This tells us that the relative contributions of the two power laws of the modified Herschel-Bulkley equation to the flow are very similar, both in bulk and at the interface. 
What changes are the values of the exponents, especially the values of $u$, which might be connected to the different confinement of the particles at the interface and/or to the fact that they cannot interact with other particles in the $z$-direction.

\section{Modeling the phase behavior at the interface}

The behavior of the microgels at the interface was modeled by means of a simple 2D effective model. 
To start with, we use the 2D Hertzian effective interaction previously established in Ref.~\cite{Cam20}, where we have calculated this explicitly between two monomer-resolved microgels at a liquid-liquid interface through Umbrella Sampling Molecular Dynamics simulations. 
The resulting potentials favourably compared to the theoretical Hertzian model in 2D, $V_H(r)$, for c=5\% up to probed distances corresponding to core-core interactions. 
This potential reads,
\begin{equation}
V_H (r) = \frac{\pi Y d^2 \left(1-r /d \right)^2}{2 \ln\left(\frac{2}{1-r/d}\right)}, \,\, 
\label{eq:2dhertzian}
\end{equation}
where $Y$ is the 2D Young modulus and $d$ is the effective diameter of the particles. 
This is the unit length of our Molecular Dynamics (MD) simulations and it is assumed to be the same as in experiments, i.e.~$d=675$~nm.  
We then perform Langevin Dynamics simulations at the same area factions as in experiments, calculated as $\zeta_{2D}=\pi d^2/4$, and varying values of $Y$ in order to identify an estimate of the Hertzian strength. All simulations are performed in the $NVT$ ensemble at constant temperature $T=\epsilon/k_B=1.0$, where $\epsilon$ is the unit of energy and $k_B$ is the Boltzmann constant. 
All particles have a unit mass $m$ and are enclosed in a two-dimensional box with periodic boundary conditions of lateral length $L$. 
The time units are expressed as $\tau=\sqrt{m \left\langle d^{2} \right\rangle /\epsilon}$. 
The time step used for the simulation is $dt=0.0005 \tau$.
We find that a good agreement with experiments is obtained for $Y=170 k_BT/d^2$, yielding a Hertzian strength $\sim 383 k_B T$.  
The Langevin friction coefficient is set to $\Gamma=10 m/\tau$.

The system undergoes a transition from fluid to hexagonal crystal at $\zeta_{2D}\sim 1.0$. 
Importantly, the position of the first peak of the $g(r)$ is in agreement with the experimental one in the range $1.0 \lesssim \zeta_{2D} \lesssim 2.75$. 
Above this packing fraction, the peak position continues to decrease to smaller and smaller distances, while in experiments a second length scale emerges.
\begin{figure}[htbp!]
    \centering
    \includegraphics[width=\textwidth]{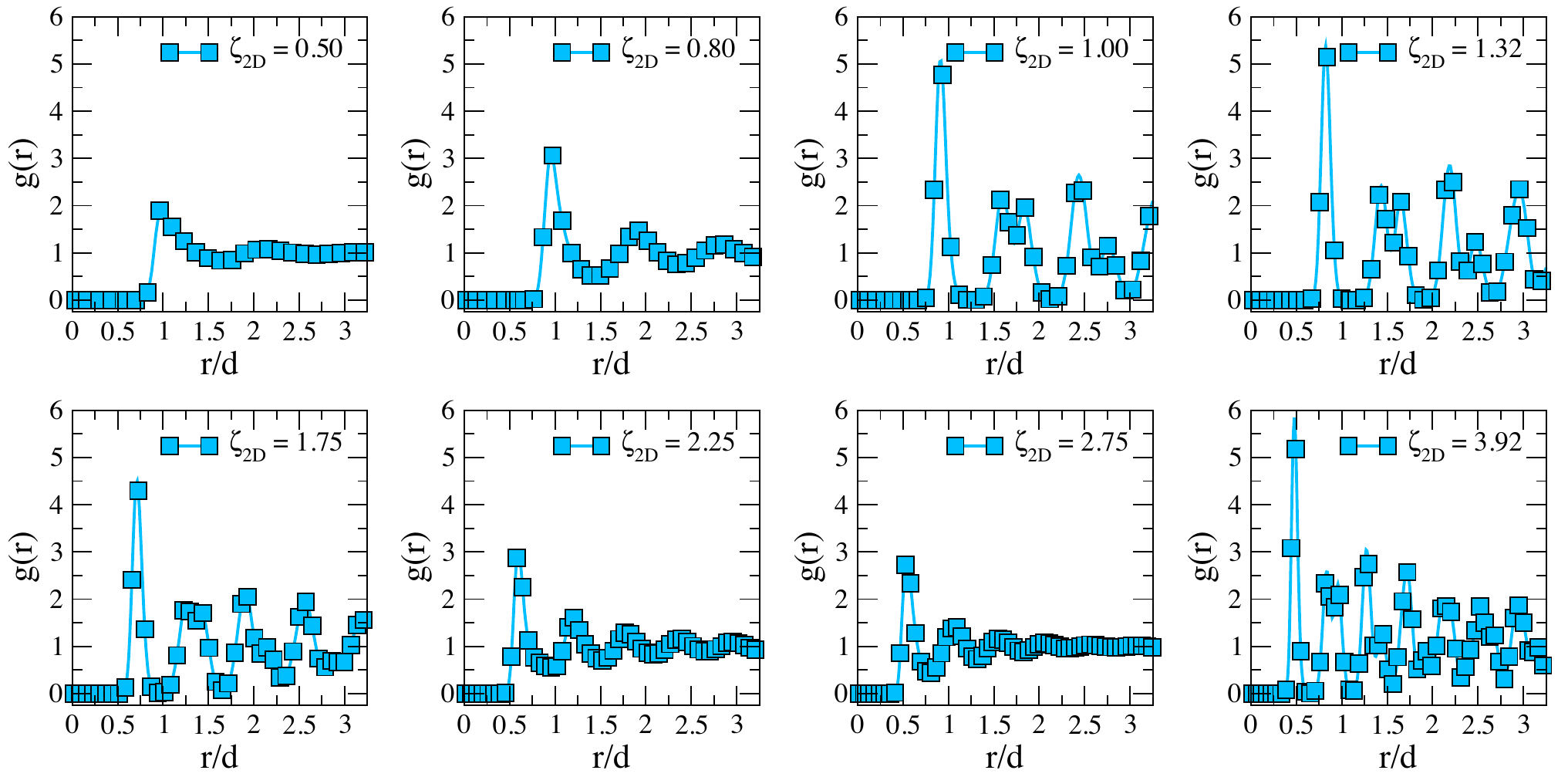}
    \caption{Radial distribution functions for the monodisperse Hertzian system. 
    For $\zeta_{2D} \gtrsim 1.0$, the system is found to be in a hexagonal lattice, which persists at all studied higher packing fractions.}
     \label{fig:gr_H}
 \end{figure}
In addition, the height of the peak in the simulations is much higher than the experimental one, due to the use of a monodisperse system. 
Indeed, the real microgels are not only polydisperse but also deformable, so that the conformation of the outer chains at the interface fluctuates in time, as a result increasing even further the effective polydispersity. 
Since this is beyond the scope of the current coarse-grained modeling, which just aims to  qualitative reproduce the experimental behavior, we use a polydisperse system with variance $s=0.06$. 
Such polydispersity in size is also reflected in the Hertzian prefactor which scales as $d^2$.
The inclusion of polydispersity only decreases the peak heights, without affecting their positions. 
This can be seen in Fig.~\ref{fig:MB5nnd}, where the value of $NND$ is the same at low $\zeta_{2D}$ for both monodisperse and polydisperse models.

However, the main point of this analysis is that the Hertzian model fails to reproduce the observed $g_{2D}(r)$ behavior at high values of packing fraction. 
This is due to the fact that a simple Hertzian model accounts only for an average elastic modulus of the particles, neglecting the fact that microgels have a harder core surrounded by a more compressible corona \cite{Sch21}. 
To account for this, we follow previous ideas that additional terms enter into play at smaller distances. 
For example, a multi-hertzian model would be appropriate to account for different elasticities, as done in \cite{bergman2018new}, but we find that, even varying the elastic modulus by more than two orders of magnitude, the second length scale does not appear. 
This indicates that a real ``core'' is at play at such small distances,  and hence we adopt a shoulder potential to model the core. 
However, we stress that we such a shoulder must necessarily be complemented by the Hertzian model at larger distances, in order to correctly capture the low and intermediate packing fraction behavior. 
We term this new potential a `square-shoulder-hertzian' (SSH) model, previously never investigated in the literature.

The resulting SSH potential, which is inspired by the continuous version of the square-shoulder put forward in Ref.~\cite{sandoval2022soft} and combined with the 2D Hertzian in Eq.~\ref{eq:2dhertzian}, is defined as,
\begin{equation}
\label{eq:VH-ss}
V_{SSH} (r) = \left\{\begin{matrix}
\frac{A_{\gamma}}{T^*} \left[\left(d/r\right)^{\gamma}-\left(d/r\right)^{\gamma-1}\right]+\frac{1}{T^*}+\epsilon_{ss}+V_H(r_0)+F_H(r_0-r) & \text{for} & r< r_{core}B_{\gamma} \nonumber \\ 
\epsilon_{ss}+V_H(r_0)+F_H(r_0-r) & \text{for} & r_{core}B_{\gamma} < r < r_0 \nonumber \\ 
\epsilon_{ss}\exp{[-K(r-r_0)^2]}+V_H(r) & \text{for} & r_0 < r < d  \nonumber\\ 
0 & \text{for} & r > d
\end{matrix}\right.
\end{equation}
\noindent where $A_{\gamma}=\gamma(\gamma/(\gamma-1.0))^{\gamma-1.0}$, $B_{\gamma}=\gamma/(\gamma-1.0)$,  $K=10000$, $\gamma=50$, $\alpha=0.354$, $T^{*}=1.474$ are taken to be the same as those validated in Ref.~\cite{sandoval2022soft}. 
Instead, the shoulder height $\epsilon_{ss}$ is fixed to be 1.0$k_BT$ and the shoulder width is fixed by the experimental values. 
Namely, $r_{core}=(215/675) d$ is the distance below which no second particle is ever found experimentally, setting a sort of hard boundary, while the ``soft'' core size is called $r_{out}=370/675 d$, estimating the core size. 
Hence, the width of the shoulder  is fixed to $r_{out}/d \sim 0.55$. 
We note incidentally that this is a much larger value than the typical ones for which iso-structural transitions are predicted in the literature for simple SS potentials~\cite{jagla1998phase,rey2017anisotropic}. 

Finally, $r_0=(r_{out}-(-log(\alpha)/K)^{1/2})\sigma$ is also fixed by Ref.~\cite{sandoval2022soft} for the current chosen width, while the Hertzian contribution is added for distances greater than $r_0$. 
To ensure the continuity of both the force and potential, additional terms are included at shorter distances, namely the constant  $V_H(r_0)$, i.e.,  the value of the Hertzian potential calculated in $r_0$, which makes the potential continuous and a linear term modulated by $F_H(r_0)$, which is the corresponding value of the Hertzian force at $r_0$, ensuring the continuity of the force. 
This last linear term included in the shoulder range makes the potential to be not completely flat, but slightly increasing towards the center of the microgel, which anyway is even more realistic than a flat profile, since the density profile of a standard microgel is usually a decreasing function of $r$, and hence, the steric repulsion should increase towards the interior of the microgel.

\begin{figure}[htbp!]
    \centering
    \includegraphics[width=\textwidth]{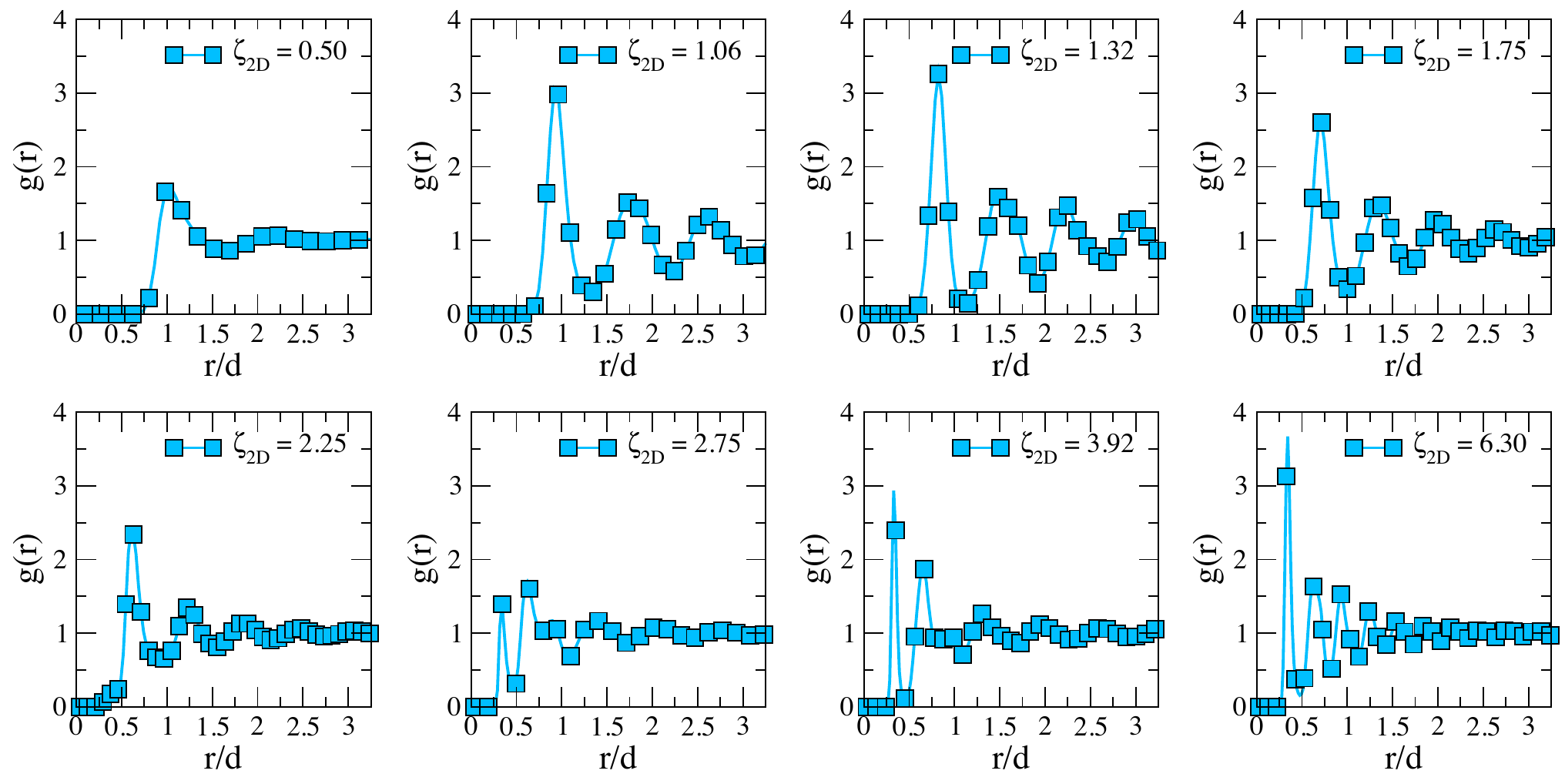}
    \caption{Radial distribution functions for the polydisperse system interacting with the Hertzian plus shoulder potential in Eq.~\ref{eq:VH-ss}.}
     \label{fig:gr_H-SS}
 \end{figure}
 
An illustration of the SSH potential is provided in the main text (Fig.~3a) and it is important to note that there are virtually no fit parameters, except for $\epsilon_{ss}$ which however only changes the relative height of the radial distribution peaks. 
All length scales are instead fixed by experiments as well as the packing fraction.

We thus perform MD simulations of the effective shoulder-Hertzian potential as a function of packing fraction, but only for the polydisperse system, amounting to $N=3229$ particles with 33 different sizes assigned following a Gaussian distribution. 
Thus, the packing fraction in the simulation is defined as $\zeta_{2D}=\frac{\pi}{4}\left\langle d^2 \right\rangle \frac{N}{A}$, where $A$ represents the area of the simulation box and $\langle d^2 \rangle $ is average squared diameter. 
The simulation protocol is identical to those used for the purely Hertzian case.

Results for the radial distribution functions are shown in Fig.~\ref{fig:gr_H-SS}, qualitatively reproducing all experimental features as a function of $\zeta_{2D}$. 
A direct comparison with experiments is provided by plotting the nearest-neighbour distance in Fig.~\ref{fig:MB5Gr}. 

In Regime I, covered by simulation data, we observe that $d_{nn}$ changes only mildly, as particles do not interact much with each other. 
After they form the hexagonal lattice, in Regime II, an isotropic shrinking, compatible with $\zeta_{2D}^{-1/2}$, is observed both in experiments and simulations. 
However, within Regime III, such isotropic shrinking is not observed any more. 
At the same time, the AFM images show microgels in closer contact and, consequent a second, much smaller nearest-neighbour distance is detected. 
This indicates a change in the interaction potential: indeed, while the behavior of the first length is well-captured by Hertzian potential only, which however continues to follow isotropic shrinking at even large packing fractions, the occurrence of a second distance is found for the Hertzian plus square shoulder one. 
This signals the onset or core-core interactions. 
Indeed, the variation of the second distance with $\zeta_{2D}$ is much smaller than for the first one, denoting a much stiffer response~\cite{Sch21, Hou22}, until in regime V the highly compressed microgels mostly behave as hard spheres and it is very likely at this point the buckling of the monolayer or the creation of multi-layers.

\begin{figure}[htbp!]
    \centering
    \includegraphics[width=0.49\textwidth]{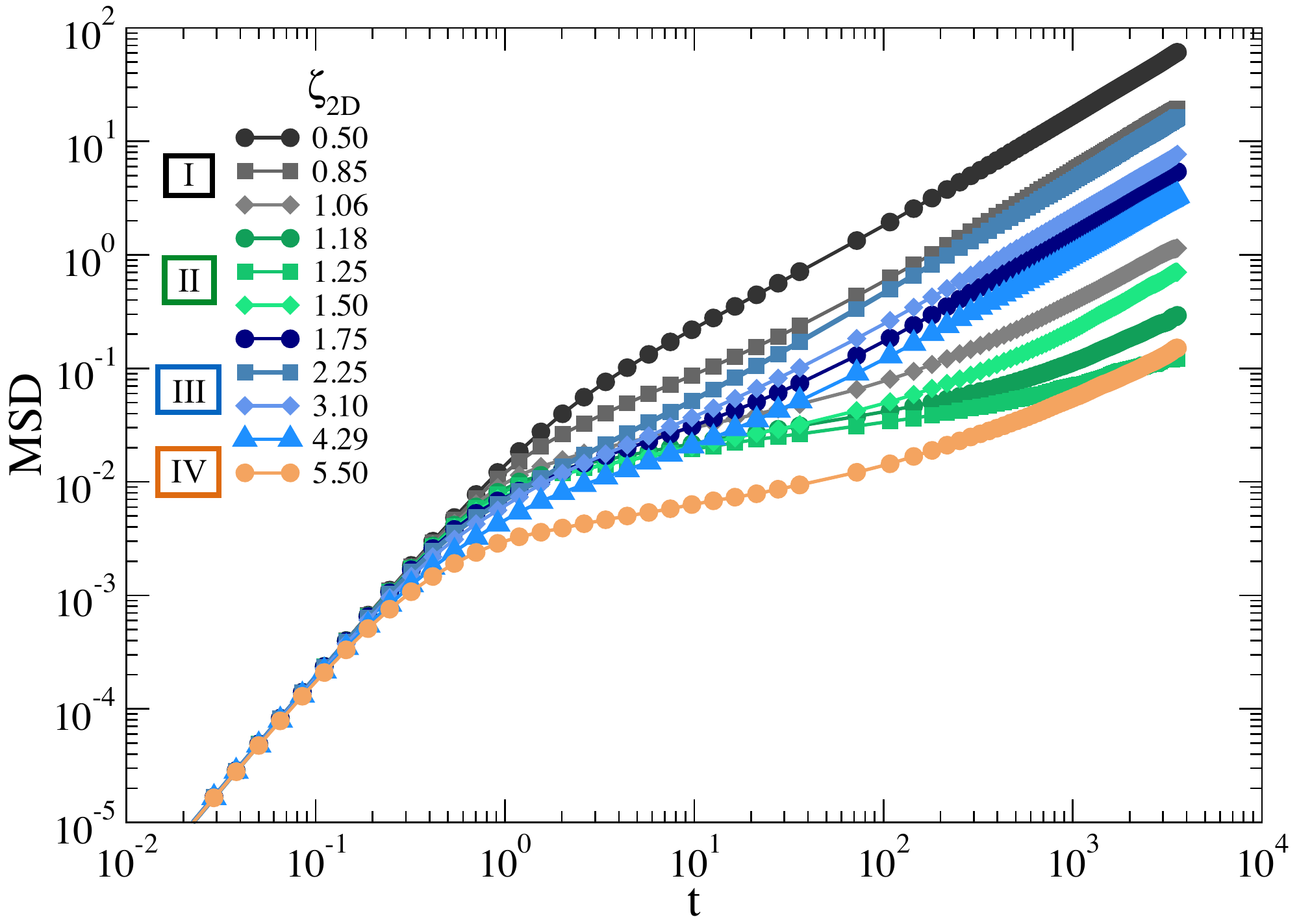}
    \caption{MSDs for the polydisperse system interacting with the Hertzian plus shoulder potential in Eq.~\ref{eq:VH-ss} as a function of time for different values of $\zeta_{2D}$. 
    The labels refer to the different regimes observed in the evolution of the moduli, that are reflected by the non-monotonic behavior of the MSD.}
     \label{fig:MSD}
 \end{figure}

For completeness, we also report in Fig.~\ref{fig:MSD} the evolution of the mean-squared displacement (MSD) with increasing area fraction. 
The plot shows that in regime I, the dynamics gets slower with increasing $\zeta_{2D}$. 
Then in regime II, in coincidence of the decrease of the elastic moduli, the dynamics gets faster and then in Region III it slows down again. 
In Region IV the system reaches a glassy state, as indicated by the much lower cage plateau of the MDS and the lack of a clear diffusive behavior at long times. 
This is why we stop at this area fraction to calculate properties, as above this value of $\zeta_{2D}$ aging and non-ergodic behavior will be present.

\section{Simulations under shear}

\begin{figure*}[htbp!]
    \centering
    \includegraphics[width=\textwidth]{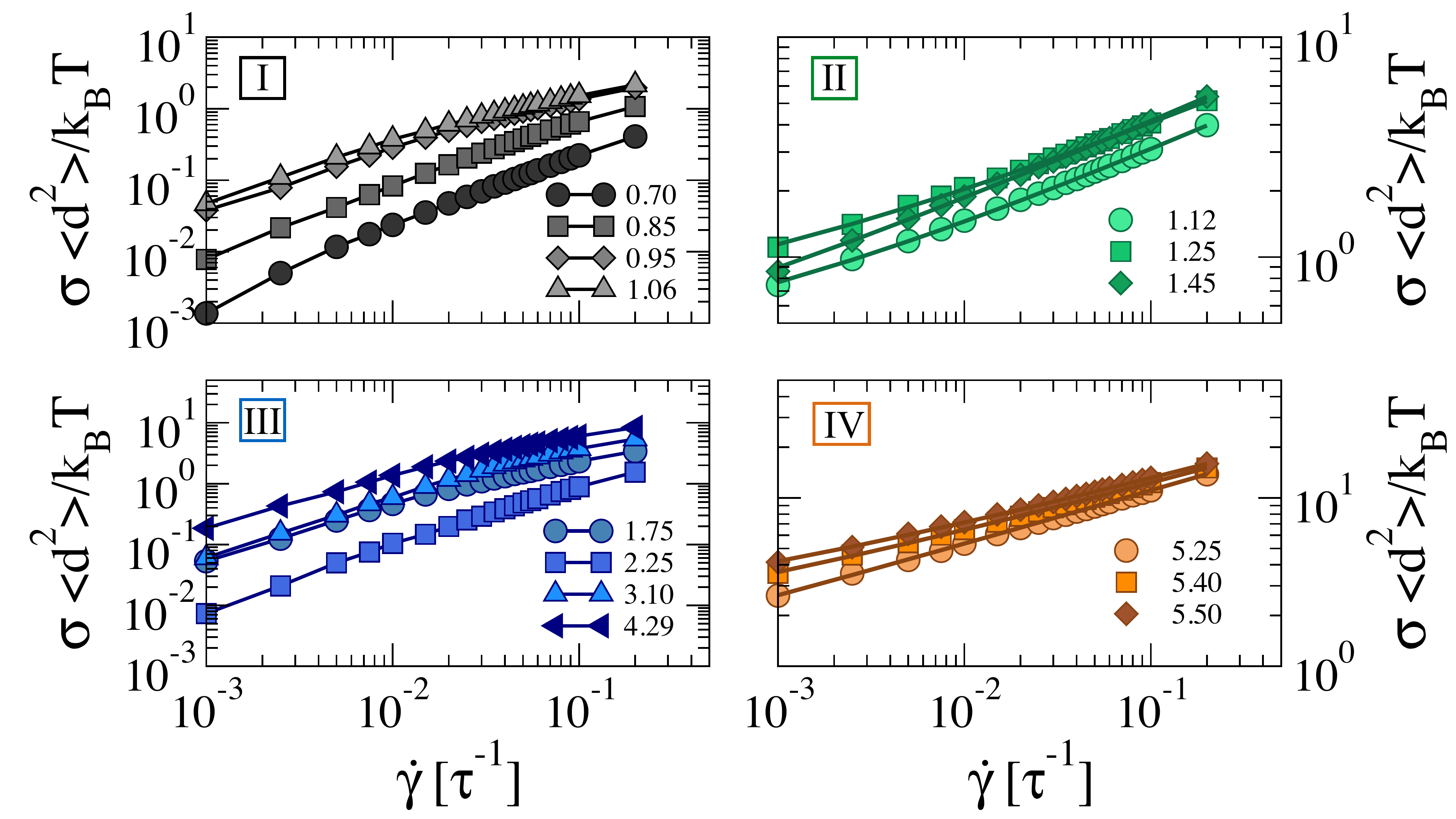}
    \caption{
     Flow curves normalized by $k_{B}T/\left\langle d^{2} \right\rangle$ as a function of the applied shear rate $\dot{\gamma}$, at different $\zeta_{2D}$. 
     Flow curves are classified in different regimens, as discussed in the main text. 
     Solid lines for data in regimes I and III are guides to the eye, whereas solid lines for data in regimes II and IV correspond to Herschel-Bulkley fits.}
    \label{Sim_FC}
\end{figure*}

To investigate numerically the rheological response of the system, we use the previously established models, namely monodisperse Hertzian and polydisperse Hertzian plus shoulder potential, and simulate them under steady shear flow with Lees-Edwards boundary conditions, following our previous work~\cite{Rui18,RuizPRM}. 
To this aim, a shear rate is imposed along the $xy$ plane in the $x-$ direction  using different shear rates $\left(\dot{\gamma}\right)$ ranging between $10^{-3}$ and $2 \times 10^{-1}$. 
During the deformation, the shear stress $\sigma_{xy}$ can be monitored using the Irving-Kirkwood expression, that is defined as:
\begin{equation}
\sigma_{xy} = \frac{1}{A} \sum_{\alpha\beta} f_{\alpha\beta}^{x} r^{y} \, ,
\label{Stress}
\end{equation}
\noindent where $f_{\alpha\beta}^{x}$ is the $x-$component of the force with respect to the pair interaction, being this either Eq.~\ref{eq:2dhertzian} or Eq.~\ref{eq:VH-ss}, $r^{y}$ corresponds the $y-$component of the distance vector between the particles $\alpha$ and $\beta$. 
At large strains $\gamma=\dot{\gamma} t$, the shear stress achieves a steady state; we compute the averaging over states sampled in steady state, $\sigma=\left\langle \sigma_{xy} \right\rangle$, to build up the flow curves represented in Fig.~\ref{Sim_FC}.

As discussed in the main text, flow curves obtained by MD simulations where the system has a solid-like response (regimes II and IV where $\sigma_y >0$) are fitted by the standard Herschel-Buckley (HB) function, $\sigma\left(\dot{\gamma} \right) = \sigma_{y} + k \dot{\gamma}^{n}$, where $\sigma_{y}$ is the estimated yield stress, $k$ is a constant and $n$ is the HB exponent. 
The fact that in experiments, there is a crossover and the need to use to two HB functions is probably not captured by the simulations because of the more reduced range in which $\dot{\gamma}$ can be varied, with respect to experiments which covers roughly 6 decades.

The obtained flow curves for the polydisperse SSH potential are reported in Fig.~\ref{Sim_FC}. 
Simulations  also cover regime I where interfacial rheology experiments cannot be performed. 
In this regime the system is liquid-like, prior to the formation of a solid hexagonal lattice. 
We then observe solid-like behavior in regime II, with the correct non-monotonic trends also found in experiments, as discussed extensively in the main text. 
Then, in regime III and IV, the simulated system becomes liquid-like again, with the HB fits giving $ \sigma_{y} =0$. 
We believe that this difference with respect to experiments can be explained by the simple coarse-grained model that we use, because in reality the coronas of the microgels will be strongly touching and interacting in this regime, but we do not capture this feature, as in our model the polymeric degrees of freedom are not retained. 
However, we correctly capture the decrease of $ \sigma_{y}$ followed by a new solid-like behavior in regime V, which agrees well with experiments. 
The latter is only found if the square-shoulder is incorporated in the model, while the Hertzian one would remain fluid-like in the whole investigated range of packing fractions. 
More details on these behaviors are discussed in the main text, where also the master scaling of the flow curves in regime II and IV/V is reported.  

\bibliographystyle{apsrev4-1}

\end{document}